\documentclass[aps,twocolumn,groupaddress,usenatbib,nofootinbib,showkeys,showpacs,altaffilletter]{revtex4-1}

\usepackage{graphicx}
\usepackage{dcolumn}
\usepackage{amssymb}
\usepackage{amsfonts}
\usepackage{amsbsy}
\usepackage{color}
\usepackage{rotating}
\usepackage[english]{babel}

\begin{document}

\title{Tension between SNeIa and BAO: current status and future forecasts}

\date{\today}

\author{Celia Escamilla-Rivera}
\email{celia\_escamilla@ehu.es}
\author{Ruth Lazkoz}
\email{ruth.lazkoz@ehu.es}
\author{Vincenzo Salzano}
\email{vincenzo.salzano@ehu.es}
\author{Irene Sendra}
\email{irene.sendra@ehu.es}
\affiliation{Fisika Teorikoaren eta Zientziaren Historia Saila, Zientzia eta Teknologia Fakultatea, \\ Euskal
Herriko Unibertsitatea, 644 Posta Kutxatila, 48080 Bilbao, Spain}

\begin{abstract}
Using real and synthetic Type Ia SNe (SNeIa)  and baryon acoustic oscillations (BAO) data representing
current observations forecasts, this paper investigates the tension between those
probes in the dark energy equation of state (EoS) reconstruction considering the
well known CPL model and Wang's low correlation reformulation. In particular, here we present
simulations of BAO data from both the the radial and
transverse directions. We also explore the influence of priors on $\Omega_m$ and
$\Omega_b$ on the tension issue, by considering $1\sigma$ deviations in either one or both
of them. Our results indicate that for some priors there is no tension between a single dataset
(either SNeIa or BAO) and their combination (SNeIa+BAO). Our criterion to discern
the existence of tension ($\sigma$-distance) is also useful to establish which is the dataset
with most constraining power; in this respect SNeIa and BAO data  switch roles when current and
future data are considered, as forecasts predict and spectacular quality improvement on BAO
data. We also find that the results on the tension are blind to the way the CPL model is
addressed: there is a perfect match between the original formulation and that by the correlation
optimized proposed in \cite{Wang:2008zh}, but the errors on the parameters are much narrower
in all cases of our exhaustive exploration, thus serving the purpose of stressing the convenience
of this reparametrization.
\end{abstract}



\maketitle

\section{Introduction}
\label{sec:Introduction}
Almost thirteen years ago, the accelerated expansion of the Universe was discovered while
reconstructing the Hubble diagram of SNeIa\cite{SNfirst}. From
then on, a large amount of cosmological data has been collected: the already mentioned
Hubble diagram of SNeIa \cite{SNsecond}; the measurements of cluster properties as the mass,
the correlation function and the evolution with redshift of their abundance \cite{cluster};
the optical surveys of large scale structure \cite{lss}; the anisotropies in the cosmic
microwave background (CMB) \cite{cmb,Komatsu10}; the cosmic shear measured from weak
lensing surveys \cite{weaklens} and the Lyman\,-\,$\alpha$ forest absorption \cite{lyman}. All
these data sets confirm that the Universe is spatially flat, exhibits a subcritical matter
content and its expansion is accelerated; but despite all these recent progresses, the origin
of this accelerated expansion remains unknown.\\

The simplest and most accepted model is the $\Lambda$CDM one, where the acceleration is driven
by a positive cosmological constant, $\Lambda$ \cite{lambda,sahni}, which has to be small
enough for it to start dominating the Universe only at late times, where it represents about
the $73\%$ of the total energy content of the Universe, as reported by the WMAP7-year
analysis \cite{wmap7matrix}. At the same time, it requires the presence of a large
amount of cold dark matter (about the $23\%$ of the total matter-energy content), non-baryonic
matter which is detectable only by its gravitational interaction with ordinary baryonic matter.
The $\Lambda$CDM model provides a good fit to most of the data \cite{lcdm}, and still remains the
best candidate in some respects, as  demonstrated by the WMAP7-year data \cite{Komatsu10}.
But it is also affected by serious theoretical shortcomings \cite{Liddle:1998ew,Chevallier01}
that have motivated the search for alternative candidates generically referred to as \textit{dark energy}.
Attempts in this direction include, among others, modified gravity
\cite{modgrav}, Cardassian cosmology \cite{Freese:2002sq}, the Chaplying gas
\cite{chap}, braneworld models \cite{branes}, and $f(R)$ theories
\cite{Sotiriou:2008rp}.

Studying the nature and the main properties of dark energy practically means studying its equation of state (EoS),
$w(a) \doteq \rho_{X}(a)/ p_{X}(a)$, where $\rho_{X}$ and $p_{X}$ are respectively its density and pressure
and $a\doteq1/(1+z)$ is the scale factor. It follows then that the dark energy density function (in critical
density units) is
\begin{equation}
\Omega_{X}(a) \propto \exp \left[3\int_{a}^{1} \frac{da'}{a'} (1+w(a')) \right] \; ,
\end{equation}
which appears in the Hubble function as
\begin{equation}
H(a) = H_{0} \left[ \Omega_{m} a^{-3}+ \Omega_{X}(a) \right]^{1/2} \; .
\end{equation}
where $\Omega_m$ is the present value of the matter density in critical density units
and $(1-\Omega_m)=\Omega_{X}(a=1)$ is the current dark energy density in the same units. Clearly, the spatial flatness hypothesis has been made, and  radiation and curvature contributions have been neglected.\\

This last equation makes it possible to study and constrain the dark energy EoS within a suitable model
and relate it to observations, although actually extra ingredients such as the fraction of baryons $\Omega_b$
or the comoving sound horizon size may be needed.  The most important and difficult goal at this stage is to
address a physically adequate expression for the dark energy EoS. To be useful,
$w(z)$ must be sufficiently sophisticated to be able to explain the data, and simple enough
so as to provide reliable predictions. Eventually, once the free parameters
in  the chosen  dark energy model and the remaining cosmological
parameters  have been observationally constrained, a picture of the evolution of the Universe will emerge.

It is quite strongly stablished that dark energy domination began somewhat recently, and therefore low redshift data,
are precisely those best suited for its analysis. The two main astrophysical tools of such nature are standard candles
(objects with well determined intrinsic luminosity) and standard rulers (objects with
well determined comoving size). Such probes provide us with distance measures related to $H(z)$, and the best so far representatives
of those two classes are SNeIa and BAO. Those are in fact low redshift datasets, and much effort  is being done in those two observational contexts toward obtaining more and better measurements \cite{Albrecht}.

It is well known, however, that a ``tension'' \cite{tensionfirsta,tensionfirstb} among the estimated EoS parameters from
different datasets in general and those two mentioned in particular can arise. The word ``tension'' will be used in this paper in its more general meaning, namely, to indicate that the EoS parameters values obtained by using a certain dataset can differ
from those obtained from another dataset at least at a $2\sigma$ level. Typically, BAO data seem to prefer phantom dark energy
($w<-1$), which in fact is the key point to be addressed, as a confirmation would definitely do away with the $\Lambda$CDM model. On the other hand, SNeIa data seem to favour a larger proportion of dark matter in the Universe than BAO data (which unlike
supernovae are both connected to the cosmological background and also to its structure). Clearly,
tension between those datasets becomes manifest in rather relevant aspects of dark energy reconstruction.

As long as tension is concerned,  one cannot waive some additional facts. First of all, priors in the fraction of dark matter and/or baryons may influence  the estimated values of the dark energy parameters rather strongly , as discussed in  \cite{tensionthird,sahni}.
Secondly, the tension may be bringing to light some caveats in the dark energy parametrization. Thirdly, it may
just be a statistical pathology produced by the limitations of the observational data available \cite{tensionthird}.

All those questions arise ,and hopefully will be given some answers, in the context
of any of the particular  parametric forms of $w(z)$ \cite{demodels} (just to cite a few) which have been
proposed to describe the behavior of time-evolving dark energy EoS. At present,  due to the current limited  constraining power of the data not one  of those parametrizations has championed definitively and there
is still a long of room for discussion about their suitability either along the focus or many other
interesting routes. Nevertheless, practicality has made some models  particularly popular \cite{Albrecht06}, with the
the Chevallier-Polarski-Linder (CPL) model \cite{Chevallier01,Linder03} excelling among them. It is given by
\begin{equation}
w(a) = w_{0} + (1-a) w_{a} \; ,
\end{equation}
where $w_{0}$ is the present value of the EoS and $w_{a}$ constraints the value of the EoS
at early times, i.e, for $a \rightarrow 0$ we have $w \sim w_{0} + w_{a}$.
This scenario has gathered many followers because it is simple and sensitive enough to good quality astronomical data.
This parametrization is consistent with low redshift data as SNeIa and BAO data,
although some unusual effects manifest themselves when CMB data are considered (in the vicinity of the regions of the
parameter space where dark energy mimicks dark matter).  Nevertheless, $w_0$ and $w_a$ are highly correlated, thus
making it harder to extract very strong conclusions about dark energy evolution from them.

A reformulation of the CPL model was proposed in \cite{Wang:2008zh} which puts the accent on the evaluation
of the same dark energy EoS at two different redshifts, one being the present one  $w_0 = w(a=1) = w(z=0)$
and the other being that corresponding to a \textit{pivot} redshift value: $w_c = w(z=z_c)$. This pivot value is
chosen appropriately so that the correlation between $w_0$ and $w_c$ is minimal, as this turns to be advantageous.

Motivated by the issue of the tension among SNeIa and BAO data, and taking into account the points already made
above in this section, we work along the following  main directions.
First, we study how the choice of priors affects the basic features of our dark energy models (separation
   from the $\Lambda$CDM model). Then, we  explore whether the influence of the choice of priors gets
   attenuated when the correlation  between parameters is minorated. After that, we repeat those tests with synthetic data based
         on the specifications of future survey properties, thus checking the influence of future systematics on EoS estimations and
errors. The mock data we present are to the best of our knowledge the first mock collections of data reproducing future observations derived from the radial and transverse BAO modes.

The paper is organized as follows: in Sec.~\ref{sec:Theory} we briefly review
the EoS parametrizations considered; in Sec.~\ref{sec:Data}~-~\ref{sec:Mock} we describe the
observational and simulated mock datasets; in Sec.~\ref{sec:Grid} we review the used statistical
methodology; in Sec.~\ref{sec:Results} we outline and comment the results of our analysis.

\section{Low correlation Dark Energy EoS}
\label{sec:Theory}
In \cite{Wang:2008zh} the popular CPL parametrization was given another turn of the screw to become
\begin{equation}\label{eq:w_wang}
w(a) = \left(\frac{a_{c}-a}{a_{c}-1}\right) w_{0} + \left(\frac{a-1}{a_{c}-1}\right) w_{c},
\end{equation}
where $w_{0} = w(z=0)$ and $w_{c} = w(z=z_{c})$. The motivation for this reformulation, as we have already pointed out and will develop further later, is to decorrelate  the parameters as much as possible. Taking into account Eq.(\ref{eq:w_wang}),
and assuming the existence of a matter component as well, the square of $E(z) = H(z)/H_0$ reads
\begin{eqnarray}\label{eq:hubble_wang}
E^2(z) &=& \Omega_m (1+z)^3 + (1-\Omega_m) (1+z)^{ 3 \left[ 1 + \small\left( \frac{a_{c} w_{0} - w_{c}}{a_{c} - 1} \right) \right]}
\nonumber \\
&\times&\exp \left\{ 3 \left(\frac{w_{c}-w_{0}}{a_{c}-1}\right) \frac{z}{1+z} \right\}.
\end{eqnarray}
Eq.(\ref{eq:w_wang}) can be related to the usual CPL parametrization by
\begin{equation}\label{eq:pivot}
w_c =w_0 +(1-a_c)w_a \, .
\end{equation}
In principle, the subindex $c$  should  indicate the scale factor (or redshift) value for which the parameters
$(w_{0},w_{c})$ are uncorrelated. However, this value depends  on the single dataset (or combination of them) one is considering; in \cite{Wang:2008zh}
the author proposed to fix it as $z_{c}=0.5$,  this value being sufficiently close to the one derived from current data
($z_{c} \sim 0.3$) and thus arguing that the correlation between $(w_{0},w_{c})$ will consequentially be relatively small. Moreover,
starting from the definition of $w_{c}$, it is straightforward to demonstrate that the couple
$(w_{0},w_{c})$ will always be less
correlated than the CPL parameter couple $(w_{0},w_{a})$, when
\begin{equation}
\sigma^2(w_0) < 2 |(1-a_{c}) \sigma^2(w_{0}w_{a})| .
\end{equation}
Note that this condition is always satisfied for $z_{c}=0.5$, i.e. $a_{c}=2/3$.
For this particular value of $a_{c}$ one gets
\begin{eqnarray}
E^2 (z) &=& \Omega_m (1+z)^3 + (1-\Omega_m) (1+z)^{3(1-2 w_0 +3 w_{0.5})} \nonumber \\
&\times& \exp{\left[\frac{9(w_0 -w_{0.5})z}{1+z}\right]} \; .
\end{eqnarray}
Before closing let us recall that we will at all stages be working with two testbenches: Wang's (low correlation) model on the one hand, and
the CPL model on the other (we omit the specifics of $E(z)$ for the CPL model as this has been reproduced in the literature
extensively).

\section{Current astrophysical data}
\label{sec:Data}

For what concerns the kind of observational data we are going to use, we remind that
the most familiar and sensitive probes for testing dark energy properties are the SNeIa: even though they are
extremely rare astrophysical events, the modern and specifically planned strategies of detection make it possible
to observe and collect them up to relatively high redshifts ($z \approx 2$) \cite{Zitrin:2011mq}, well beyond the era when dark energy
should have become the dominant energy term in the Universe. Another important property of modern surveys is that
they offer a statistically significant number of observations, whose quality
has improved considerably over the years \cite{Sanchez:2009ka}.
Present samples are made of approximately 500 events; future surveys are planned to detect up to approximately 5000 events.
However, these SNeIa datasets are not always consistent with other types of cosmological observations (as we shall see),
and even an historical tension can be detected among different SNeIa samples \cite{Sanchez:2009ka,tensionfirsta}.

While SNeIa are well known to be standard candles, we also have standard rulers; among these,
one of the main techniques rests on the BAO peaks
detection in the galaxy power spectrum. This has recently emerged as a promising standard ruler for
cosmology, potentially enabling precise measurements of the dark energy parameters with a
minimum of systematic errors \cite{Wang:2010gq,Blake:2005jd}. For this technique to be
useful, one must explore, as uniformly
as possible, a large volume in real space with a sufficiently high sampling density. These requirements are met by current high quality data (the Sloan Digital Sky Survey (SDSS) \cite{York00,Abazajian09} and
the 2-degree Field Galaxy Redshift Survey (2dFGRS) \cite{Colless03}), and thus BAO have become a nature tool
to study the cosmic expansion history.

\subsection{Supernovae}
\label{sec:SNdata}

We use the most recent SneIa sample available, the Union2
sample described in \cite{Amanullah10}. The Union2 SNeIa
compilation is the result of a new dataset of low-redshift nearby-Hubble-flow
Type Ia SNe and is built with new analysis procedures to work with several
heterogeneous SNeIa compilations. It includes the Union data set
from \cite{Kowalski08} with six added SNeIa first presented in
\cite{Amanullah10}, along with SNeIa from \cite{Amanullah08}, the low-z
and the intermediate-z data from \cite{Hicken09a} and
\cite{Holtzman08} respectively. After the application of various selection cuts
to create a homogeneous and high signal-to-noise
data set, $\mathcal{N}_{\mathrm{SN}}=557$ SNeIa events distributed over the
redshift interval $0.015 \leq z \leq 1.4$ were obtained.

The statistical analysis of the Union2 SNeIa sample rests on the definition
of the modulus distance:
\begin{equation}
\mu(z_{j}, \mu_0) = 5 \log_{10} [ d_{L}(z_{j}, \Omega_m; \boldsymbol{\theta}) ] + \mu_0,
\end{equation}
where $d_{L}(z_{j}, \Omega_m; \boldsymbol{\theta})$ is the Hubble free luminosity
distance:
\begin{equation}\label{eq:dl_H}
d_{L}(z, \Omega_m; \boldsymbol{\theta}) = (1+z) \ \int_{0}^{z} \mathrm{d}z'
\frac{1}{E(z', \Omega_m; \boldsymbol{\theta})} \; .
\end{equation}
With the chosen notation we clarify the different roles of the various cosmological
parameters appearing in the formulae: the matter density parameter $\Omega_m$ appears separated as it is assumed to be fixed
to a prior value, while $\boldsymbol{\theta}$ is the EoS parameters vector ($\boldsymbol{\theta}=(w_{0},
w_{0.5})$ for Wang's model and $\boldsymbol{\theta}=(w_{0},
w_{a})$ for the CPL one), which are the parameters that we will be constraining. The best fits will be obtained by minimizing the
quantity
\begin{equation}\label{eq: sn_chi}
\chi^{2}_{\mathrm{SN}}(\mu_{0}, \boldsymbol{\theta}) = \sum^{\mathcal{N}_{\mathrm{SN}}}_{j =
1} \frac{(\mu(z_{j}, \Omega_m; \mu_{0}, \boldsymbol{\theta})\} -
\mu_{obs}(z_{j}))^{2}}{\sigma^{2}_{\mathrm{\mu},j}},
\end{equation}
where the $\sigma^{2}_{\mathrm{\mu},j}$ are the measurement
variances.
The nuisance parameter $\mu_{0}$ encodes the Hubble
parameter and the absolute magnitude $M$, and has to be
marginalized over. Giving the heterogeneous origin of the Union2 dataset, and the procedures to
reduce data described in \cite{Kowalski08}, we will work with an alternative version \cite{altchi} of
Eq.~(\ref{eq: sn_chi}), which consists in minimizing the quantity
\begin{equation}\label{eq: sn_chi_mod}
\tilde{\chi}^{2}_{\mathrm{SN}}(\boldsymbol{\theta}) = c_{1} -
\frac{c^{2}_{2}}{c_{3}}
\end{equation}
with respect to the other parameters. Here
\begin{equation}
c_{1} = \sum^{\mathcal{N}_{\mathrm{SN}}}_{j = 1} \frac{(\mu(z_{j}, \Omega_m ; \mu_{0}=0,
\boldsymbol{\theta})\} -
\mu_{obs}(z_{j}))^{2}}{\sigma^{2}_{\mathrm{\mu},j}}\, ,
\end{equation}
\begin{equation}
c_{2} = \sum^{\mathcal{N}_{\mathrm{SN}}}_{j = 1} \frac{(\mu(z_{j}, \Omega_m; \mu_{0}=0,
\boldsymbol{\theta})\} -
\mu_{obs}(z_{j}))}{\sigma^{2}_{\mathrm{\mu},j}}\, ,
\end{equation}
\begin{equation}
c_{3} = \sum^{\mathcal{N}_{\mathrm{SN}}}_{j = 1}
\frac{1}{\sigma^{2}_{\mathrm{\mu},j}}\,.
\end{equation}
It is easy to see that $\tilde{\chi}^{2}_{SN}$ is just a
version of $\chi^{2}_{SN}$, minimized with respect to $\mu_{0}$.
To that end, it suffices to notice that
\begin{equation}
\chi^{2}_{\mathrm{SN}}(\mu_{0}, \boldsymbol{\theta}) = c_{1} - 2 c_{2}
\mu_{0} + c_{3} \mu^{2}_{0} \; ,
\end{equation}
which clearly becomes minimum for $\mu_{0} = c_{2}/c_{3}$, and so
we can see $\tilde{\chi}^{2}_{\mathrm{SN}} \equiv
\chi^{2}_{\mathrm{SN}}(\mu_{0} = 0, \boldsymbol{\theta})$. Furthermore,
one can check that the difference between $\chi^{2}_{SN}$ and
$\tilde{\chi}^{2}_{SN}$ is negligible.

\subsection{Baryon Acoustic Oscillations}
\label{sec:BAOdata}

BAO have rarely been used independently of other datasets (such as SNeIa or the CMB) due to the small
number of datapoints available so far \cite{baofirst}, but they provide a
measuring stick for a better understanding of the nature of the accelerated expansion
of the Universe. Moreover, they stand on a completely independent basis, as compared to the SNeIa technique,
and can contribute important features by comparing the data of the sound horizon today (using the
galaxies clustering) to the sound horizon at the time of recombination (extracted from the CMB).

In \cite{Percival10} the authors analyze the clustering of galaxies within the
spectroscopic Sloan Digital Sky Survey (SDSS) Data Release 7 (DR7) galaxy
sample, including both the Luminous Red Galaxy (LRG) and Main samples, and
also the 2-degree Field Galaxy Redshift Survey (2dFGRS) data. In total,
the sample comprises $893319$ galaxies over $9100$ deg$^2$. For a redshift survey in a thin
shell, the position of the BAO peak approximately constrains the ratio
\begin{equation}
d_{z} \equiv \frac{r_{s}(z_{d})}{D_{V}(z)} \; ,
\end{equation}
where $r_{s}(z_{d})$ is the comoving sound horizon at the baryon dragging epoch
\begin{equation}
r_{s}(z_{d}) = \frac{c}{H_{0}} \int_{z_{d}}^{\infty} \frac{c_{s}(z)}{E(z)} \mathrm{d}z\; ,
\end{equation}
with $c$ the light velocity, $c_{s}$ the sound speed and $z_{d}$ the dragging epoch redshift. By definition,
the dilation scale $D_{V}(z)$ is
\begin{equation}
D_{V}(z,\Omega_m; \boldsymbol{\theta}) = \left[ (1+z)^2 D_{A}^2 \frac{c \, z}{H(z, \Omega_m; \boldsymbol{\theta})} \right]^{1/3}
,
\end{equation}
where $D_{A}$ is the angular diameter distance
\begin{equation}
D_{A}(z,\Omega_m; \boldsymbol{\theta}) = \frac{1}{1+z} \int_{0}^{z} \frac{c \, \mathrm{d}z'}{H(z', \Omega_m; \boldsymbol{\theta})} \; .
\end{equation}
Again in this case we will choose $\Omega_m$ to be fixed to the chosen prior value, while $\boldsymbol{\theta}$
will be the EoS parameters vector. Through the comoving sound horizon, the distance ratio $d_{z}$ is related to the expansion parameter $h$
(defined such that $H \doteq 100 h$) and the physical densities $\Omega_{m}$ and $\Omega_{b}$. Specifically, following  \cite{Percival10,Komatsu09} we have
\begin{equation}
r_{s}(z_{d}) = 153.5 \left( \frac{\Omega_{b} h^2}{0.02273}\right)^{-0.134} \left( \frac{\Omega_{m} h^2}{0.1326}\right)^{-0.255} \; \mathrm{Mpc} \; .
\end{equation}
Modeling the comoving distance-redshift relation as a cubic spline in the parameter $D_{V}(z)$, in
\cite{Percival10} the authors obtained the following best fit results:
\begin{eqnarray}
d_{0.2}  &=& 0.1905 \pm 0.0061 \, , \nonumber \\
d_{0.35} &=& 0.1097 \pm 0.0036 \, ,
\end{eqnarray}
with an inverse covariance matrix
\begin{eqnarray}
\mathbf{C}^{-1}_{\mathbf{BAO}}=\left(\begin{array}{cc}
30124  &  -17227	\\
-17227  &  86977 \\
\end{array} \right) \; .
\end{eqnarray}

The $\chi^2$ function for BAO is defined as
\begin{equation}\label{chibao}
\chi^2_{\mathrm{BAO}}(\boldsymbol{\theta}) = \mathbf{X}^T_{\mathbf{BAO}}(\boldsymbol{\theta}) \mathbf{C}^{-1}_{\mathbf{BAO}} \mathbf{X}_{\mathbf{BAO}}(\boldsymbol{\theta}),
\end{equation}
where $\mathbf{X}_{\mathbf{BAO}}$ is defined as
\begin{equation}
\mathbf{X_{BAO}}=\left(\begin{array}{c}
\frac{r_s (z_d)}{D_V (0.2,\Omega_m;\boldsymbol{\theta})}    -0.1905	\\
\frac{r_s (z_d)}{D_V (0.35,\Omega_m;\boldsymbol{\theta})}    -0.1097 \\
\end{array} \right).
\end{equation}

\section{Mock data}
\label{sec:Mock}

Even though some astrophysical data like supernovae have a considerable quality and samples are made of
a significant number of datapoints, the derived constraints on dark energy are still not narrow enough, and
therefore uncertainties on its nature and features are still considerable.
This is particularly important as regards BAO observations, which is becoming a very active area with many planned
future surveys \cite{Albrecht}. Thus, forecasting the improvements
to be expected from future missions is of paramount interest. The argued relevance of the questions we propose in this paper
suggests the usefulness of reexamining them in the light of synthetic data generated under the specifications
of future surverys. Our main tool for this task is the free software for cosmological calculations
Initiative for Cosmology (iCosmo) (see \cite{Icosmo}) and its 
BAO modules.  This package is the ofspring of considerable theoretical efforts in different forecast aspects
\cite{Blake:2005jd}, and we  have modified and extended it to suit it to our needs.

\subsection{SNeIa simulations}
\label{sec:SNmock}

To create SNeIa mock samples we have taken into account the specifications of the
Wide-Field Infrared Survey Telescope (WFIRST)\footnote{http://wfirst.gsfc.nasa.gov/}, a space mission planned by NASA which has among its
primary objective is to explore the nature of dark energy employing three distinct techniques, measurements of BAO,
SNeIa distances, and weak gravitational lensing.
In \cite{WFIRST} and Table~\ref{tab:WFIRST} we report the  redshift distribution of the expected observed SNeIa
within various redshift bins.
The main redshift range extends up to $z\sim 1.6$, with $4$ expected SNeIa at $z>1.6$. The  very
low redshift subsample, i.e. the $500$ SNeIa at $z<0.1$, is supposed to come from the Nearby Supernova Factory
(SNfactory) \cite{SNfactory} with measurements in the redshift range $[0.03,0.08]$ so we are going
to assume the former as the lowest redshift value in our mock sample.

Our formula for errors on SNeIa magnitudes stems from a recipe used in the binning approach \cite{snerror},
which we have adapted to the case with one supernovae per redshift bin:
\begin{equation}\label{eq:error}
\sigma_{m}^{\mathit{eff}} = \sqrt{{\sigma_{int}^2}+ \sigma_{pec}^2 + \sigma_{syst}^2} \; ,
\end{equation}
where 
\begin{itemize}
 \item $\sigma_{int} = 0.15$ is the intrinsic dispersion in magnitude per SN,
       assumed to be constant and independent of redshift for all  well-designed surveys;
 \item $\sigma_{pec} = {5 \sigma_{v}}/{(\ln(10) c z)}$ is the error due to the uncertainty
       in the SN peculiar velocity, with $\sigma_{v} = 500$ km$/$s, $c$ is the velocity of light  and $z$ is the redshift for any SN;
 \item $\sigma_{syst} = 0.02 ({z}/{z_{max}})$ is the floor uncertainty related to all the irreducible
       systematic errors with cannot be reduced statistically by increasing the number of observations.
       The value $0.02$ is conservative from the perspective of what space-based missions could achieve. Those
       are precisely the resources expected to provide high-redshift SN, which are in turn the ones
       in which the systematic error term is expected to contribute. Note as well that $z_{max}$ is the maximum 
       observable redshift in
       the considered mission and this linear term in redshift is used to account for
       the dependence with redshift of many of the possible systematic error sources (for example the
       Malmquist bias or gravitational lensing effects).
\end{itemize}
In addition, we have included some extra (though very slight) noise, and then we have checked that our mock data
are compliant with the main features and trends of the Union2 sample.

For our analysis we have use three different fiducial models, all derived from the WMAP7-year analysis
\cite{wcdm}: one is the quiessence model derived by using only CMB data and is named CMB-oriented model,
another one is the quiessence model coming out from combining CMB data with BAO and is named BAO-oriented model, and
the last one is the quiessence model coming out from combining CMB data with SNeIa and is named SNeIa-oriented model. The main
difference among them is the value of the EoS parameter, $w$: in the CMB-oriented case it clearly corresponds to  a phantom model, i.e.
$w<-1$; in the BAO-oriented model the regime is phantom (though only slightly),  whereas in the SNeIa-oriented model $w>-1$.

The $\chi^2$ function for the SNeIa mock data has been constructed as described in  Sec.~(\ref{sec:SNdata}).
The supernovae redshift distribution we have follow is tabulated in (following \cite{Albrecht}); then, through a simple extension to 
iCosmo of our own we managed to generate the $d_L$ values for those redshifts as drawn from hundreds of universes randomly and normally distributed around the fiducial one, and finally a reduction was been performed to give our mock sample. We again produced
three mock samples using the previously mentioned fiducial scenarios.

The use of different datasets offers a wider picture of the problem and allows for more robustness in our conclusions, which on 
the other are only indicative and not as substantial as those coming from real data.
{\renewcommand{\tabcolsep}{2.mm}
{\renewcommand{\arraystretch}{1.25}
\begin{table}
\begin{minipage}{0.4\textwidth}
\caption{Redshift distribution of WFIRST SNeIa samples}\label{tab:WFIRST}
\resizebox*{0.6\textwidth}{!}{
\begin{tabular}{cc}
\hline
\hline
{Redshift bin}  &  {$\#$ of SNeIa}  \\
\hline
\hline
$<0.1$     & $500$  \\
$0.1-0.2$  & $200$  \\
$0.2-0.3$  & $320$  \\
$0.3-0.4$  & $445$  \\
$0.4-0.5$  & $580$  \\
$0.5-0.6$  & $660$  \\
$0.6-0.7$  & $700$  \\
$0.7-0.8$  & $670$  \\
$0.8-0.9$  & $110$  \\
$0.9-1.0$  & $80$   \\
$1.0-1.1$  & $25$   \\
$1.1-1.2$  & $16$   \\
$1.2-1.3$  & $16$   \\
$1.3-1.4$  & $4$    \\
$1.4-1.5$  & $4$    \\
$1.5-1.6$  & $4$    \\
$>1.6$     & $4$    \\
\hline
\hline
\end{tabular}}
\end{minipage}
\end{table}}}

\subsection{BAO simulations}
\label{sec:BAOmock}

Future BAO surveys will target at larger volumes and improved statistical features, so the result will be narrower constraints
on dark energy. The current approach (as described in Sec.~\ref{sec:BAOdata}) considers the quantity $D_{V}$, which is (modulo
some constants) the geometric mean of the following two quantities which are becoming the new observational targets
\begin{equation}
y(z) \equiv \frac{r(z)}{r_{s}(z_{r})} \qquad \mathrm{and} \qquad y'(z) \equiv \frac{r'(z)}{r_{s}(z_{r})} \; ,
\end{equation}
where
\begin{itemize}
  \item $r(z)$ is the comoving distance to redshift slice $z$ and it is related to the transverse
        BAO mode;
  \item $r'(z) \equiv \mathrm{d}r(z) / \mathrm{d}z = c/(H_{0}E(z))$
        is the derivative of $r(z)$, and it is related to the
        radial BAO mode; and
  \item $r_{s}(z_{r})$ is the sound horizon at recombination.
\end{itemize}
In general, BAO measurements are more sensitive to dark energy evolution as they involve $1/H(z)$ directly, whereas luminosity  data
from SNeIa are related to an integral of that quantity so sensitivity will typically be somewhat compromised.

\begin{figure}[htbp]
\centering
\includegraphics[width=8.0cm]{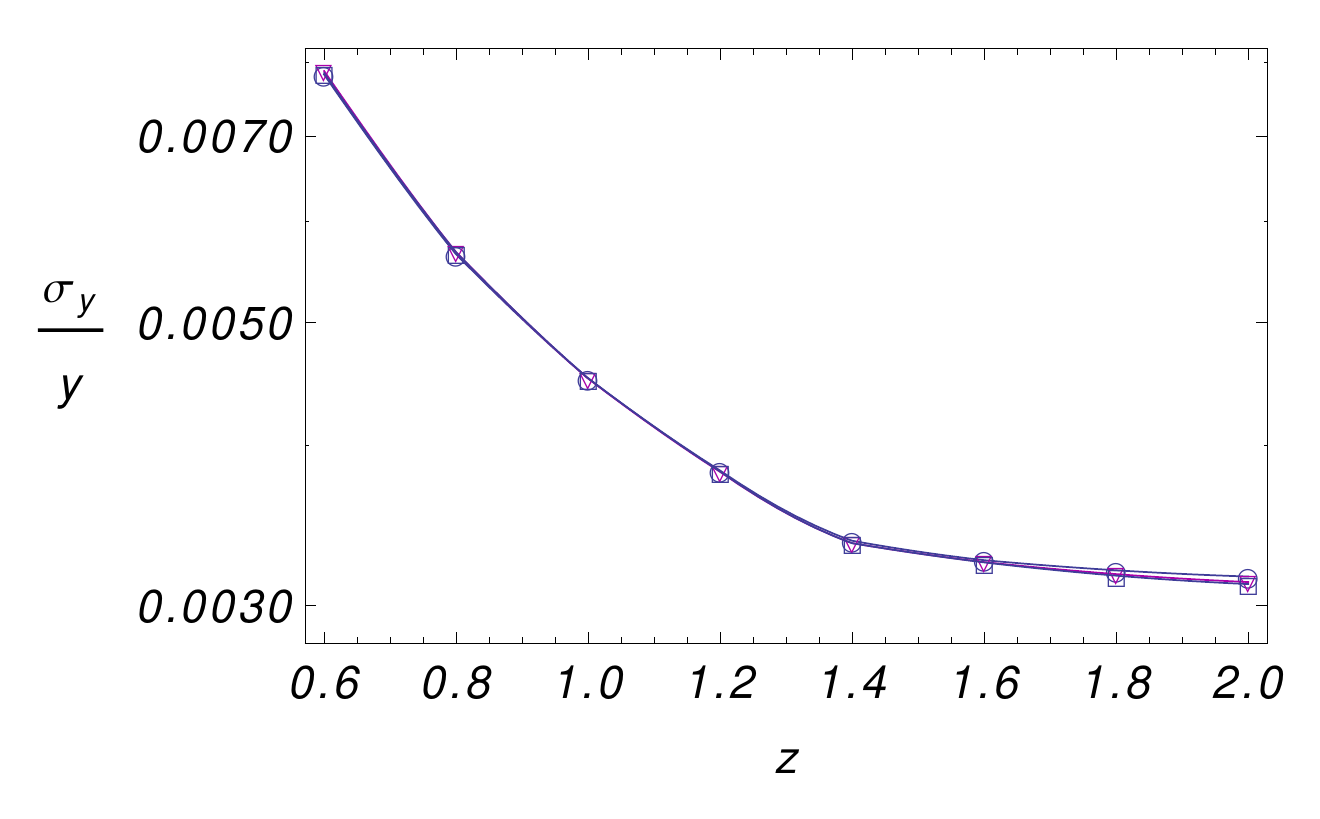}
\includegraphics[width=8.0cm]{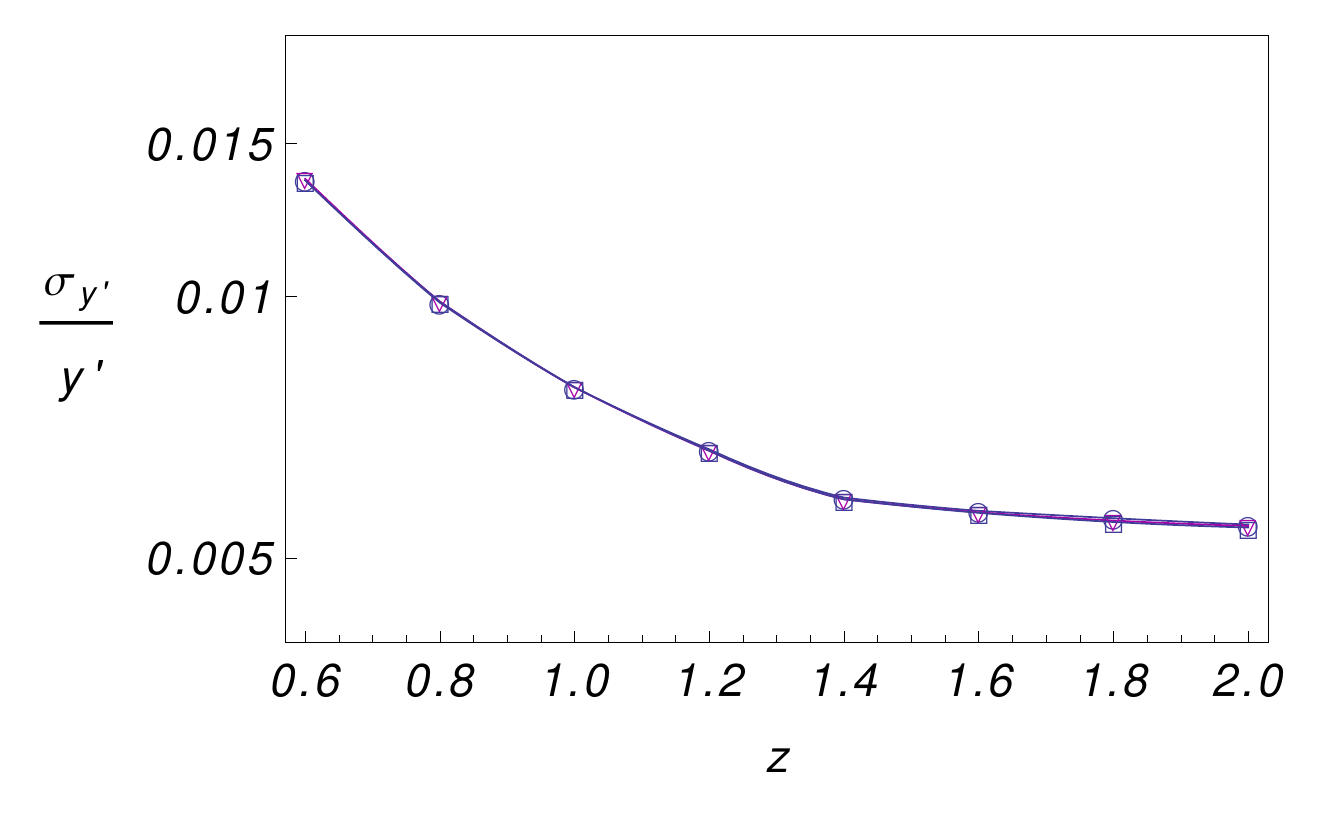}
\caption{BAO mock data: fractional errors of BAO quantities $y(z)$ and $y'(z)$ for the Euclid survey.}
\label{fig:BAO-mock1}
\end{figure}

The publicly available code has built-in routines based on the universal BAO fitting
formulae for the diagonal errors on $y$ and $y'$ presented in
\cite{Blake:2005jd}.
 We have made the proper modifications of the code to replace the defaults  with 
 the EUCLID \cite{euclid} survey properties as specified in: this survey is expected to cover $20000$ sq. deg. and observe
$6.1\times10^7$ galaxies in the redshift range $0.5<z<2.1$, and  as it  is a spectroscopic survey we fix the redshift precision
as $0.001(1+z)$. Then,  
 we have
 written extra codes to generate 
 a large number of normal random realizations around the mentioned three fiducial models 
 Finally, after some reduction, the synthetic BAO data presented in Tables
in Tables ~\ref{tab:baosim1}~-~\ref{tab:baosim2}~-~\ref{tab:baosim3} and Fig.~(\ref{fig:BAO-mock1}) have been obtained.


To define the $\chi^2$ function for BAO mock data we have to take into account
that $y$ and $y'$ are correlated: such correlation can be quantified by a correlation coefficient
$\rho \equiv {\sigma_{yy'}}/{\sigma_{y}\sigma_{y'}} \approx 0.4$ \cite{SeoEisenstein07}. With this
in mind, the $\chi^2$ function is now defined as \cite{Coe09}
\begin{eqnarray}
\chi^{2}_{\mathrm{BAO}}(\boldsymbol{\theta}) &=& \frac{1}{1-\rho^2} \left( \sum^{\mathcal{N}_{\mathrm{BAO}}}_{j =
1} \frac{(y(z_{j}, \Omega_m, \Omega_b; \boldsymbol{\theta}) -
y(z_{j}))^{2}}{\sigma^{2}_{y,j}} \right. \nonumber \\
&+& \left. \frac{(y'(z_{j}, \Omega_m, \Omega_b; \boldsymbol{\theta}) -
y'(z_{j}))^{2}}{\sigma^{2}_{y',j}} \right. \nonumber \\
&-& \left. 2 \rho  \frac{y(z_{j}, \Omega_m, \Omega_b; \boldsymbol{\theta}) -
y(z_{j})}{\sigma_{y,j}}\times \right. \nonumber \\
&& \left. \frac{y'(z_{j}, \Omega_m, \Omega_b; \boldsymbol{\theta}) -
y'(z_{j})}{\sigma_{y',j}}+
\right) \, ,
\end{eqnarray}
which clearly reduces to the more familiar expression for $\chi^2$ when
the observable quantities are uncorrelated (i.e. $\rho = 0$).

{\renewcommand{\tabcolsep}{1.75mm}
{\renewcommand{\arraystretch}{1.25}
\begin{table}
\begin{minipage}{0.5\textwidth}
\caption{Simulated BAO data for $y$ and $y'$: CMB-oriented sample.}
\resizebox*{0.6\textwidth}{!}{
\begin{tabular}{ccccc}
\hline \hline
$z$ & $y$ & $\sigma_{y}$ & $y'$ & $\sigma_{y'}$\\
\hline
$0.6$ & $15.005$ & $0.117$ & $21.401$ & $0.291$\\
$0.8$ & $19.221$ & $0.109$ & $18.953$ & $0.186$\\
$1.0$ & $22.708$ & $0.102$ & $17.012$ & $0.137$\\
$1.2$ & $25.762$ & $0.098$ & $15.098$ & $0.101$\\
$1.4$ & $28.635$ & $0.096$ & $13.421$ & $0.079$\\
$1.6$ & $31.126$ & $0.101$ & $12.213$ & $0.069$\\
$1.8$ & $33.540$ & $0.107$ & $10.987$ & $0.061$\\
$2.0$ & $35.357$ & $0.112$ & $10.094$ & $0.055$\\
\hline
\hline
\end{tabular}\label{tab:baosim1}}
\end{minipage}
\end{table}}}

{\renewcommand{\tabcolsep}{1.75mm}
{\renewcommand{\arraystretch}{1.25}
\begin{table}
\begin{minipage}{0.5\textwidth}
\caption{Simulated BAO data for $y$ and $y'$: BAO-oriented sample.}
\resizebox*{0.6\textwidth}{!}{
\begin{tabular}{ccccc}
\hline \hline
$z$ & $y$ & $\sigma_{y}$ & $y'$ & $\sigma_{y'}$\\
\hline
$0.6$ & $15.021$ & $0.118$ & $21.488$ & $0.294$\\
$0.8$ & $19.094$ & $0.108$ & $19.147$ & $0.189$\\
$1.0$ & $22.768$ & $0.103$ & $17.049$ & $0.134$\\
$1.2$ & $25.910$ & $0.099$ & $15.216$ & $0.101$\\
$1.4$ & $28.751$ & $0.096$ & $13.672$ & $0.080$\\
$1.6$ & $31.354$ & $0.102$ & $12.340$ & $0.070$\\
$1.8$ & $33.725$ & $0.107$ & $11.130$ & $0.061$\\
$2.0$ & $35.840$ & $0.112$ & $10.154$ & $0.055$\\
\hline
\hline
\end{tabular}\label{tab:baosim2}}
\end{minipage}
\end{table}}}

{\renewcommand{\tabcolsep}{1.75mm}
{\renewcommand{\arraystretch}{1.25}
\begin{table}
\begin{minipage}{0.5\textwidth}
\caption{Simulated BAO data for $y$ and $y'$: SNeIa-oriented sample.}
\resizebox*{0.6\textwidth}{!}{
\begin{tabular}{ccccc}
\hline \hline
$z$ & $y$ & $\sigma_{y}$ & $y'$ & $\sigma_{y'}$\\
\hline
$0.6$ & $14.859$ & $0.116$ & $20.881$ & $0.284$\\
$0.8$ & $18.770$ & $0.106$ & $18.624$ & $0.184$\\
$1.0$ & $22.300$ & $0.101$ & $16.619$ & $0.131$\\
$1.2$ & $25.442$ & $0.097$ & $14.854$ & $0.099$\\
$1.4$ & $28.296$ & $0.095$ & $13.364$ & $0.078$\\
$1.6$ & $30.810$ & $0.100$ & $12.089$ & $0.068$\\
$1.8$ & $33.131$ & $0.105$ & $10.939$ & $0.060$\\
$2.0$ & $35.181$ & $0.110$ & $9.983$ & $0.054$\\
\hline
\hline
\end{tabular}\label{tab:baosim3}}
\end{minipage}
\end{table}}}

\section{Two-dimensional grid analysis}
\label{sec:Grid}

Given the appropriate $\chi^2(\boldsymbol{\theta})$ function we then proceed to find the values of the
model parameters which minimize it. Using a grid (following
prescriptions in \cite{numerical}) we study the  parameter space of the model searching for the $\chi^2$ minimum.
It is well known that when the number of parameters gets large, grid techniques require an increasing computational and time consuming effort, so it becomes preferable to use
 other statistical methods, such as Markov Chain Monte Carlo (MCMC).
As we are going to explore two dark energy parameters  only (with $\Omega_{m}$ and $\Omega_{b}$
fixed), the grid method is suitable for our purposes.

Combining some intuitive guesses with the knowledge provided by the vaste literature on the topic, we typically
require just two grids (the second being finer and narrower) to find a suitable initial point to
feed the Levenberg-Marquadt algorithm (described in full detail in \cite{numerical}, pp.$~683~-~685$), and eventually detect
the minimum $\chi^2$ value. In addition, the algorithm calculates the \textit{estimated} covariance matrix $\mathbf{C}$ for the model parameters $\boldsymbol{\theta}$, which we use to
plot the confidence regions in the 2-dimensional parameter space under the Gaussian approximation. This means that provided
\begin{equation}
\Delta \chi^2(\boldsymbol{\theta}) = (\boldsymbol{\theta} - \boldsymbol{\theta_{\mathbf{best}}})^T \mathbf{C}^{-1} (\boldsymbol{\theta} - \boldsymbol{\theta_{\mathbf{best}}}),
\end{equation}
the  $1\sigma$ ($68.3\%$)  and $2\sigma$ ($95.4\%$ confidence levels) will correspond to
$\Delta \chi^2 = 2.30$ and $\Delta \chi^2 = 6.17$
respectively.

Finally, to compare results and test the issues of interest (in particular the tension among
datasets), we compute the so called $\sigma$-distance, $d_{\sigma}$, i.e. the distance in units of $\sigma$
between the best fit points of the total sample and the best fit points of (respectively) SNeIa, BAO and a $\Lambda$CDM model
corresponding to the different ($\Omega_m$,$\Omega_b$) pairs that define our priors.
Following \cite{numerical}, the $\sigma$-distance is calculated by solving
\begin{equation}
1- \Gamma(1,\vert\Delta \chi_{\sigma}^2/2\vert)/\Gamma(1) = \mathrm{erf}(d_{\sigma}/\sqrt{2}).
\end{equation}

For homogeneity and consistency our ``ruler'' is in every case the total
$\chi^2$ function ($\chi^2_{tot} = \chi^2_{\mathrm {SN}}+\chi^{2}_{\mathrm {BAO}}$), and our recipe is the following:
if we want to calculate the tension between SNeIa and SNeIa+BAO and the best fit parameter vectors
are respectively $\boldsymbol{\theta}_{\mathbf{SN}}$ and $\boldsymbol{\theta}_{\mathbf{SN+BAO}}$ then
the previous $\Delta \chi_{\sigma}^2$ will be defined as $\chi_{tot}^2(\boldsymbol{\theta}_{\mathbf{SN+BAO}}) - \chi_{tot}^2(\boldsymbol{\theta}_{\mathbf{SN}})$. Other cases follow simply from this prescription.

{\renewcommand{\tabcolsep}{2.mm}
{\renewcommand{\arraystretch}{1.5}
\begin{table}
\begin{minipage}{0.5\textwidth}
\caption{Planck priors and CPL - Wang parametrization. Minimum correlation redshift $z_c$.}
\resizebox*{0.75\textwidth}{!}{
\begin{tabular}{cccc}
\hline \hline
\multicolumn{1}{c}{$\mathrm{Priors}$} & \multicolumn{3}{c}{$z_c \; \mathrm{(CPL/Wang)}$} \\
\hline \hline
$(\Omega_{m}, \Omega_{b})$ & SNeIa & BAO & SNeIa+BAO \\
\hline
$(0.214,0.0444)$ & $0.226$ & $0.113$ & $0.220$ \\
$(0.237,0.0444)$ & $0.222$ & $0.112$ & $0.215$ \\
$(0.260,0.0444)$ & $0.217$ & $0.112$ & $0.210$ \\
$(0.214,0.0405)$ & $0.227$ & $0.113$ & $0.220$ \\
$(0.237,0.0405)$ & $0.222$ & $0.112$ & $0.216$ \\
$(0.260,0.0405)$ & $0.217$ & $0.111$ & $0.209$ \\
$(0.214,0.0366)$ & $0.226$ & $0.112$ & $0.221$ \\
$(0.237,0.0366)$ & $0.222$ & $0.112$ & $0.216$ \\
$(0.260,0.0366)$ & $0.217$ & $0.111$ & $0.210$ \\
\hline
\hline
\end{tabular}\label{tab:pivot}}
\end{minipage}
\end{table}}}

\section{Results and Conclusions}
\label{sec:Results}

As we have discussed in Sec.~\ref{sec:Introduction}, there are arguments suggesting that  considering dark energy
constraints from the combination of SNeIa and BAO data only is both relevant and useful, and so is comparing
the individual predictions drawn from each other.
This does not mean that we are going to  avoid completely possible clues from the CMB
analysis; in particular, the chosen priors for the cosmological density parameters $\Omega_m$ and $\Omega_b$
are derived from a forecast of CMB observations with the Planck satellite
mission as reported in  \cite{Burigana10}. The predicted best fits and $1\sigma$ errors are $\Omega_m h^2 = 0.1308 \pm 0.0008$
and $\Omega_b h^2 = 0.0223$ which, combined with our choice for $h = 0.742 \pm 0.036$ from \cite{Riess09},
become $\Omega_m = 0.237 \pm 0.023$ and $\Omega_b = 0.0405 \pm 0.0039$. We then consider those  best fits and combine them
 with their $1\sigma$ errors errors to produce nine
$(\Omega_m, \Omega_b)$ couples. We then feed these priors into our pipeline
so that we can describe quite in detail how the estimated EoS parameters depend on the uncertainties in $(\Omega_m, \Omega_b)$. Final results
for current data are shown in Tables ~(\ref{tab:CPL})~-~(\ref{tab:Wang}), respectively for the CPL and Wang models; whereas
forecasts for synthetic data can be examined by looking at Tables.~(\ref{tab:CPL_mock})~-~(\ref{tab:Wang_mock}).

Let us remind that Wang's EoS model is based on the idea of obtaining a minimum correlation
among its parameters. The pivot value $w_{0.5}$ is a conservative choice put forward in \cite{Wang:2008zh}, which
achieves quite a low degree of correlation and provides simple expression, although each single dataset or
combination considered will have its own optimal correlation redshift which typically will not be too far away from Wang's
choice. The specific value of this pivot redshift can be easily calculated applying error propagation
to the definition
of $w_c$ in Eq.~(\ref{eq:pivot}), and then computing the correlation parameters between $w_0$ and $w_c$
as a function of $a_c$ (or $z_c$) which one finally minimizes. For the CPL model the final expression
is \cite{Wang:2008zh}
\begin{equation}
a_c = 1+ \frac{\sigma^2(w_0)}{\sigma^2(w_0 w_a)},
\end{equation}
which clearly depends on the data set considered. For real data (see Tab.~(\ref{tab:CPL})), we effectively we detect a mean value of $z_c \approx 0.22-0.23$
using our chosen SNeIa data set, which gets only slightly reduced to $z_c \approx 0.21-0.22$ when considering the total sample SNeIa+BAO,
and $z_c \approx 0.11$ for BAO data. 

\subsection{Tension: results}
\label{sec:Tension_Res}

The tension is confirmed both in the CPL and in the Wang case, and both with current and mock data; but it also shows
different features with respect to references \cite{tensionfirsta,tensionfirstb}.

\subsubsection{Current data}
\begin{figure*}
\centering
\includegraphics[width=18cm]{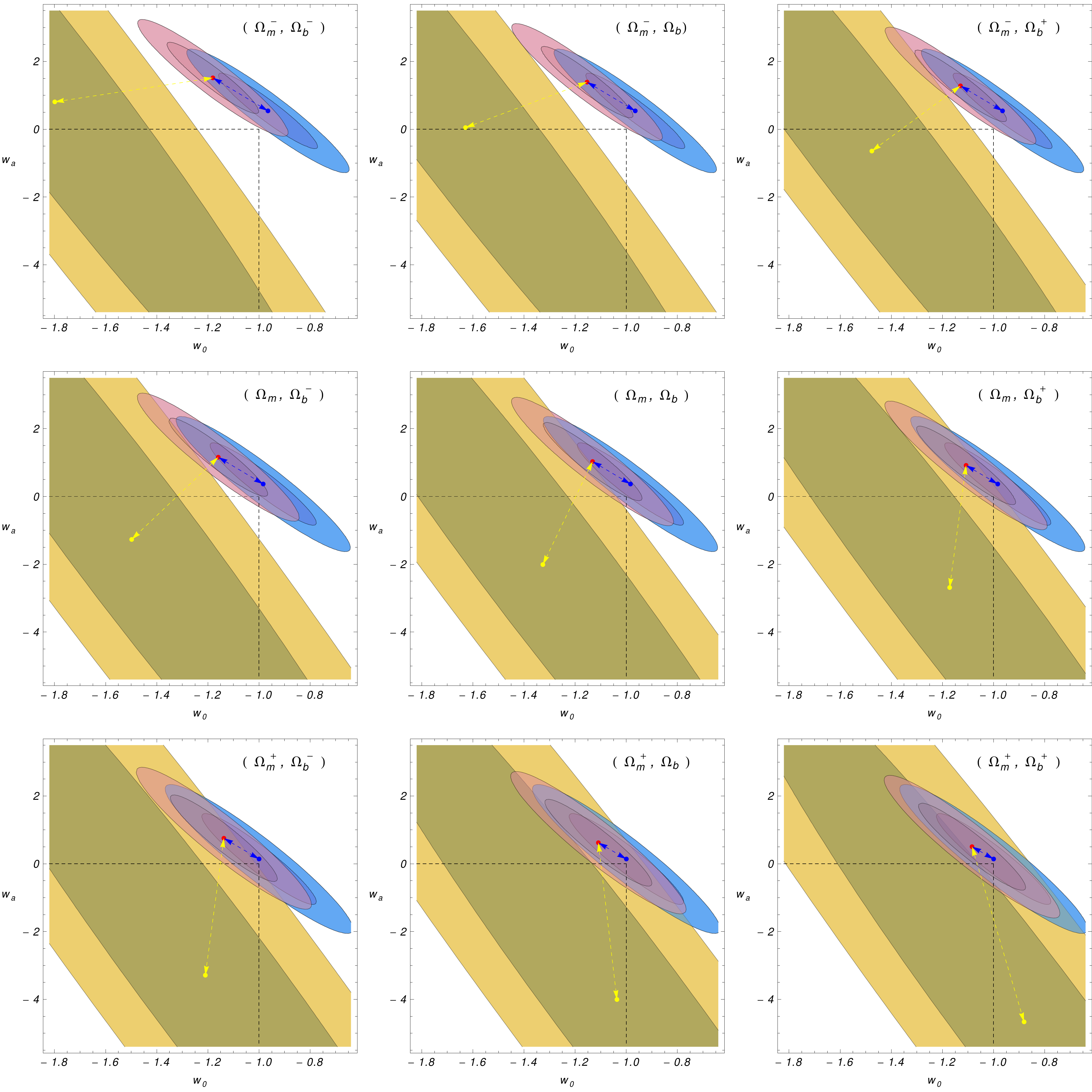}
\caption{CPL model: 1$\sigma$ and 2$\sigma$ contours in the $w_0$-$w_a$ parameter space.
In all plots SNeIa are represented by the blue region/point, BAO are the yellow region/point and the combined
SNeIa-BAO are represented by the red region/point. The point where the dashed lines cross indicates the $\Lambda$CDM model. The
arrows indicate the $\sigma$-distances for SNeIa and BAO, with the total $\chi^2$ as measure.}
\label{fig:CPL}
\end{figure*}

\begin{figure*}
\centering
\includegraphics[width=18cm]{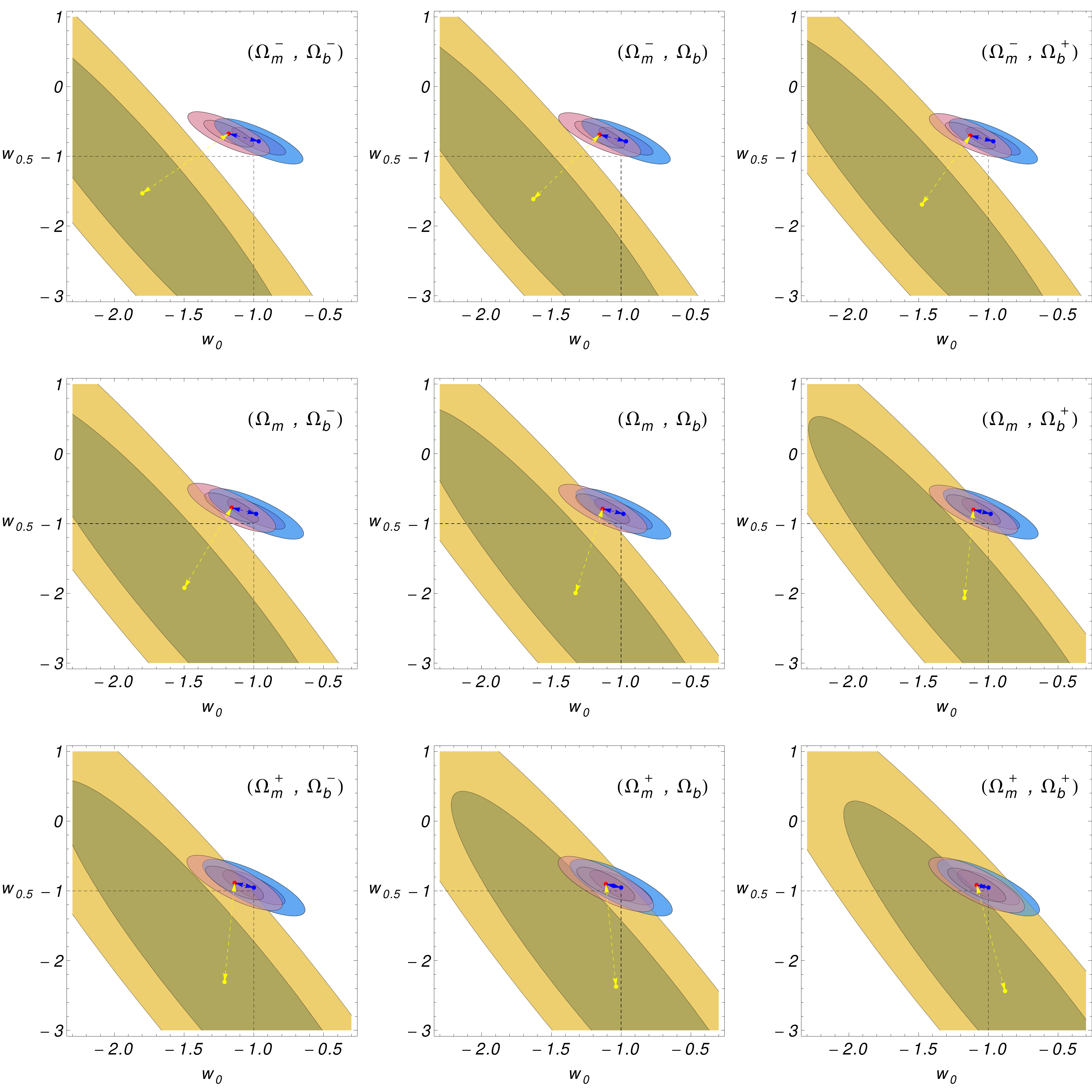}
\caption{Wang model: 1$\sigma$ and 2$\sigma$ contours in the $w_0$-$w_{0.5}$ parameter space.
In all plots SNeIa are represented by the blue region/point, BAO are the yellow region/point and the combined
SNeIa-BAO are represented by the red region/point. The point where the dashed lines cross indicates the $\Lambda$CDM model. The
arrows indicate the $\sigma$-distances for SNeIa and BAO, with the total $\chi^2$ as measure.}
\label{fig:Wang}
\end{figure*}

In these cases, SNeIa and BAO best fits can differ by more than $2\sigma$, as
it is possible to conclude just by visual inspection from Figs.~(\ref{fig:CPL})~-~(\ref{fig:Wang}). When we compare the best fits from SNeIa alone and the SNeIa+BAO combination
case we conclude there is no tension, as in all cases we have a separation smaller than
$2\sigma$. On the contrary, there is always tension between BAO alone  and SNeIa+BAO. Resorting again to visual examination
we infer that for current data, the SNeIa+BAO credible intervals are very much influenced both in position and size by the
SNeIa data, as these have much more statistical weight.

We can also underline another important property: the tension depends  strongly on the value of the cosmological priors.
In particular, we find a precise trend: the $\sigma$-distance between SNeIa and SNeIa+BAO best fits gets reduced either when  $\Omega_m$ or
$\Omega_b$ grow. In other words,
compliance between SNeIa and BAO data improves when the amount of dark energy decreases, as  has been noticed in
earlier works \cite{biggerom}. If $\Omega_m$ and
$\Omega_b$ are large enough ($\Omega_m=0.260$ and $\Omega_b=0.0444,0.0405$),  we even get close encounters (i.e. the distance among the best fits is less
than $1\sigma$). The same trend holds true between SNeIa and BAO in the sense that the larger either  $\Omega_m$ or
$\Omega_b$ become, the larger the overlap between credible intervals.

As expected, there is large sensitivity to changes in $\Omega_m$ (from which both SNeIa and BAO depend) as
compared to variations in $\Omega_b$ (which
only influences BAO aspects).  Note that BAO contours  are  much larger in size
thus reflecting their statistically less meaningful character. 

\subsubsection{Mock future data}

\begin{figure*}
\centering
\includegraphics[width=18cm]{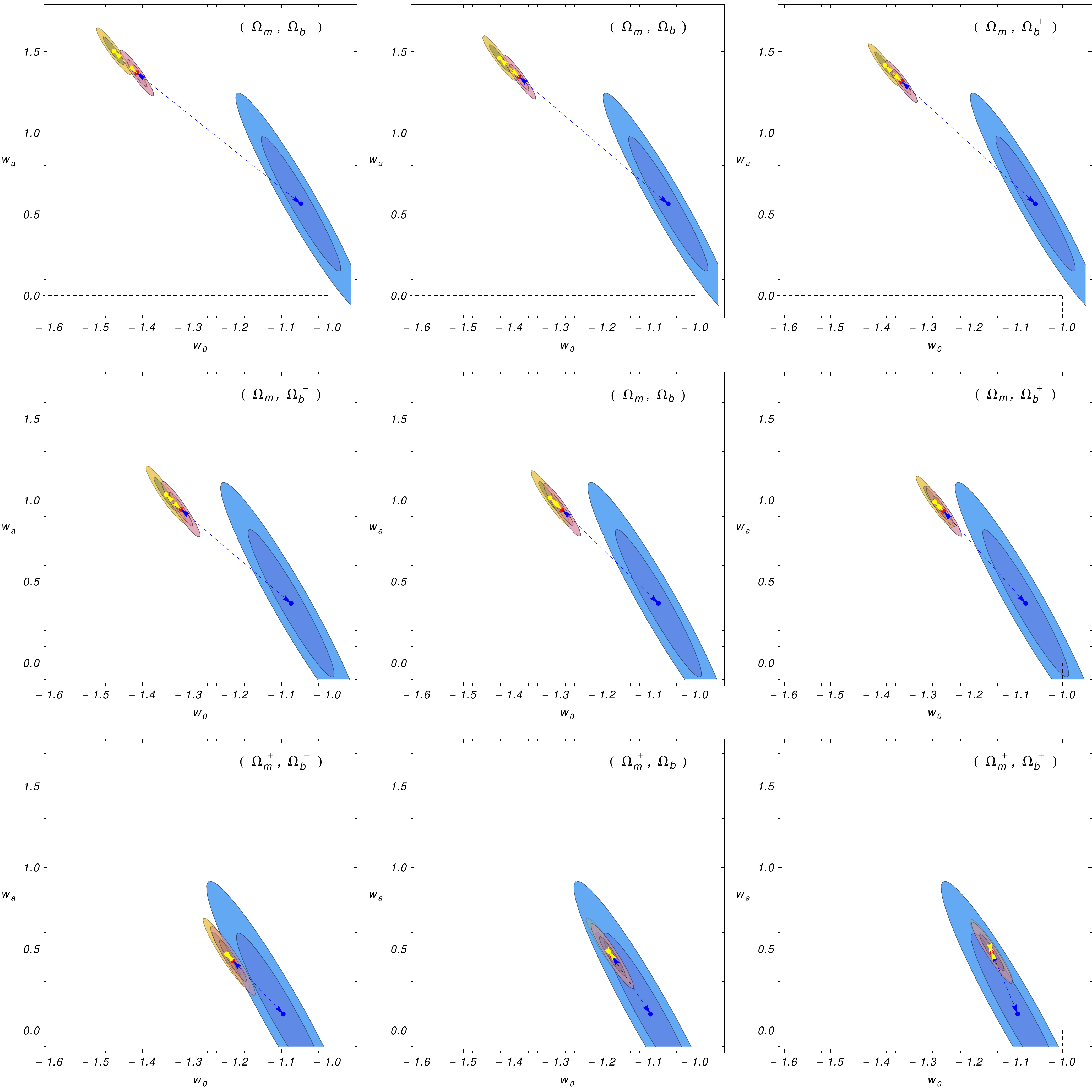}
\caption{CPL model with CMB-oriented mock data: 1$\sigma$ and 2$\sigma$ contours in the $w_0$-$w_a$ parameter space.
In all plots SNeIa are represented by the blue region/point, BAO are the yellow region/point and the combined
SNeIa-BAO are represented by the red region/point. The point where the dashed lines cross indicates the $\Lambda$CDM model. The
arrows indicate the $\sigma$-distances for SNeIa and BAO, with the total $\chi^2$ as measure.}
\label{fig:CPL_WMAP}
\end{figure*}

\begin{figure*}
\centering
\includegraphics[width=18cm]{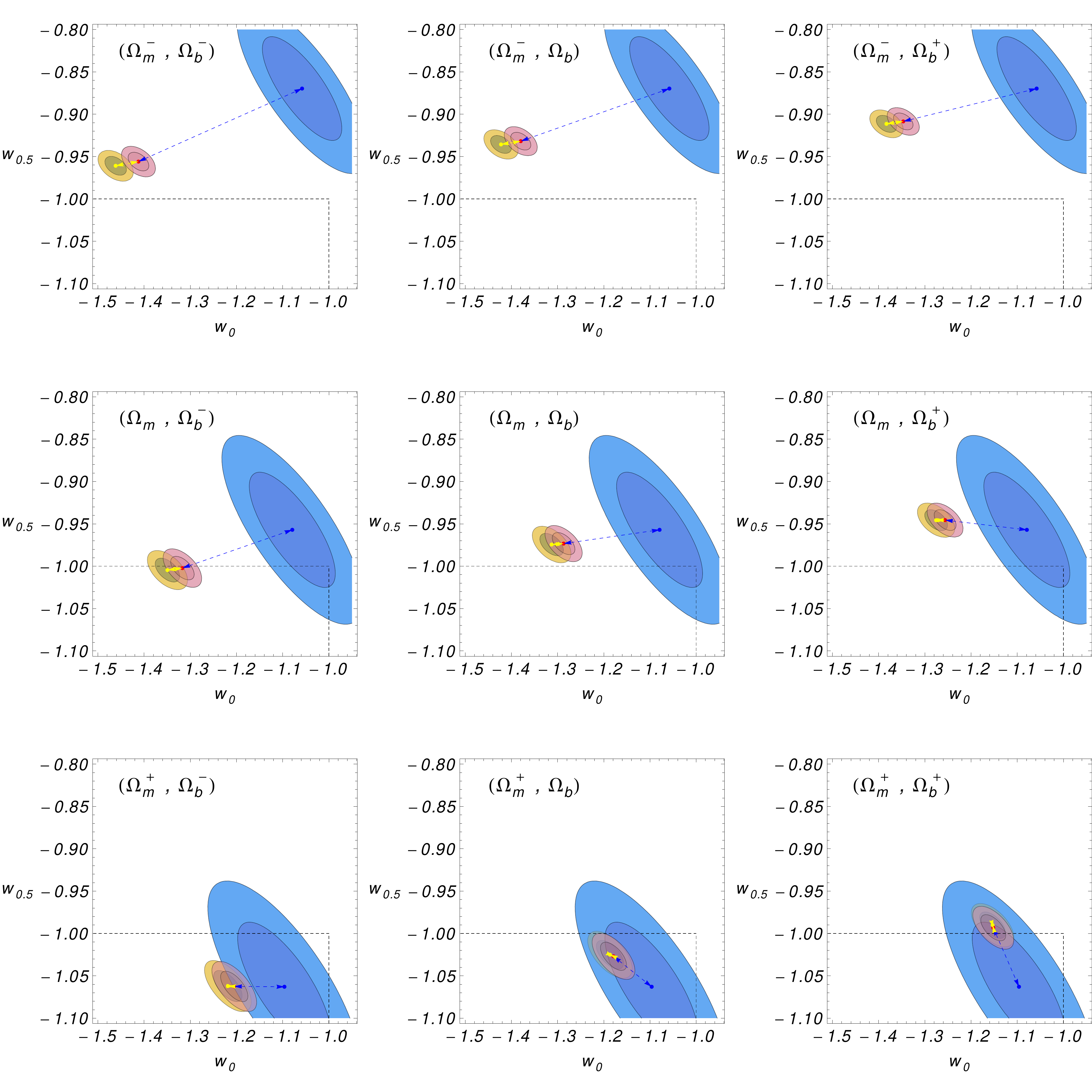}
\caption{Wang model with CMB-oriented mock data: 1$\sigma$ and 2$\sigma$ contours in the $w_0$-$w_{0.5}$ parameter space.
In all plots SNeIa are represented by the blue region/point, BAO are the yellow region/point and the combined
SNeIa-BAO are represented by the red region/point. The point where the dashed lines cross indicates the $\Lambda$CDM model. The
arrows indicate the $\sigma$-distances for SNeIa and BAO, with the total $\chi^2$ as measure.}
\label{fig:Wang_WMAP}
\end{figure*}

\begin{figure*}
\centering
\includegraphics[width=18cm]{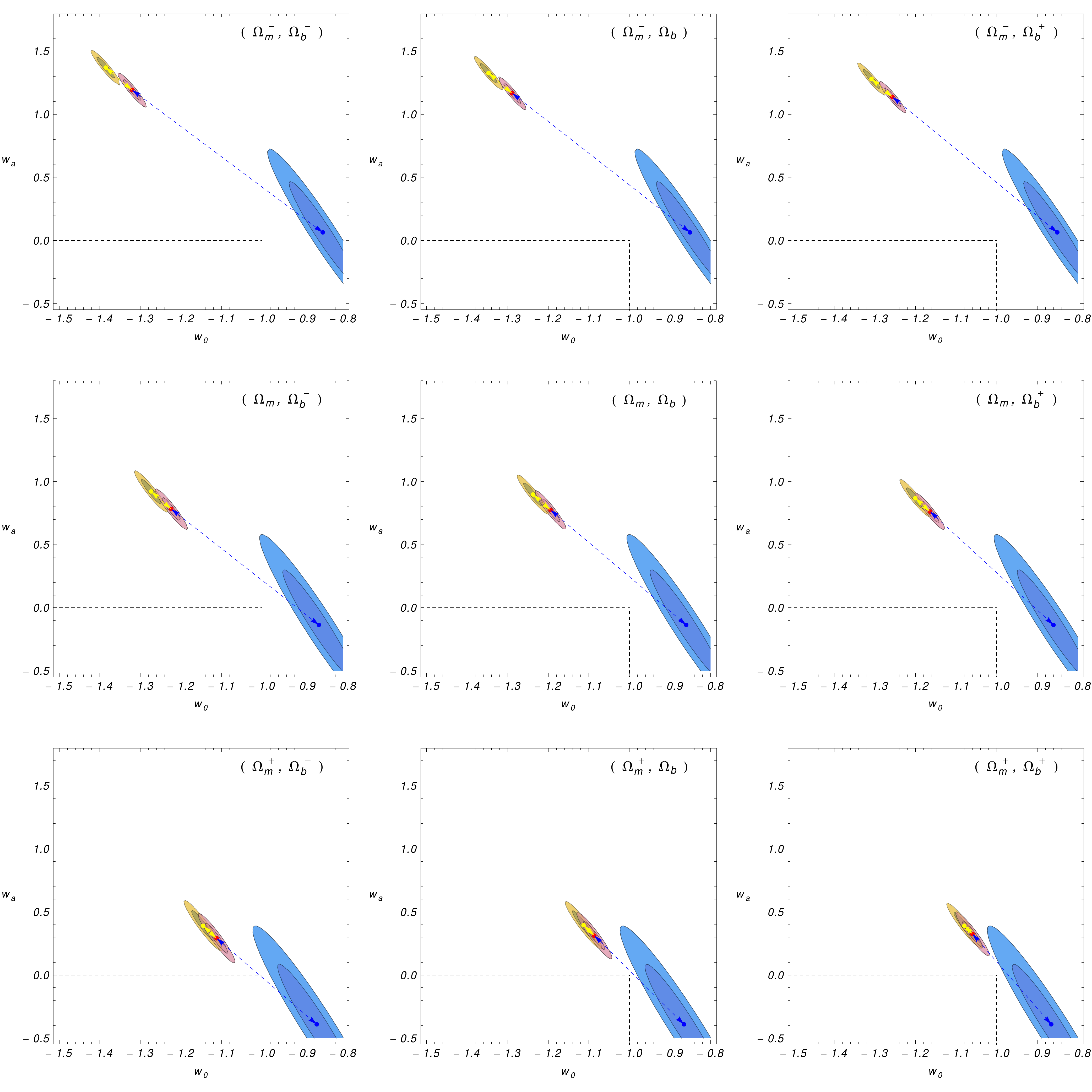}
\caption{CPL model with SNeIa-oriented mock data: 1$\sigma$ and 2$\sigma$ contours in the $w_0$-$w_a$ parameter space.
In all plots SNeIa are represented by the blue region/point, BAO are the yellow region/point and the combined
SNeIa-BAO are represented by the red region/point. The point where the dashed lines cross indicates the $\Lambda$CDM model. The
arrows indicate the $\sigma$-distances for SNeIa and BAO, with the total $\chi^2$ as measure.}
\label{fig:CPL_SN}
\end{figure*}

\begin{figure*}
\centering
\includegraphics[width=18cm]{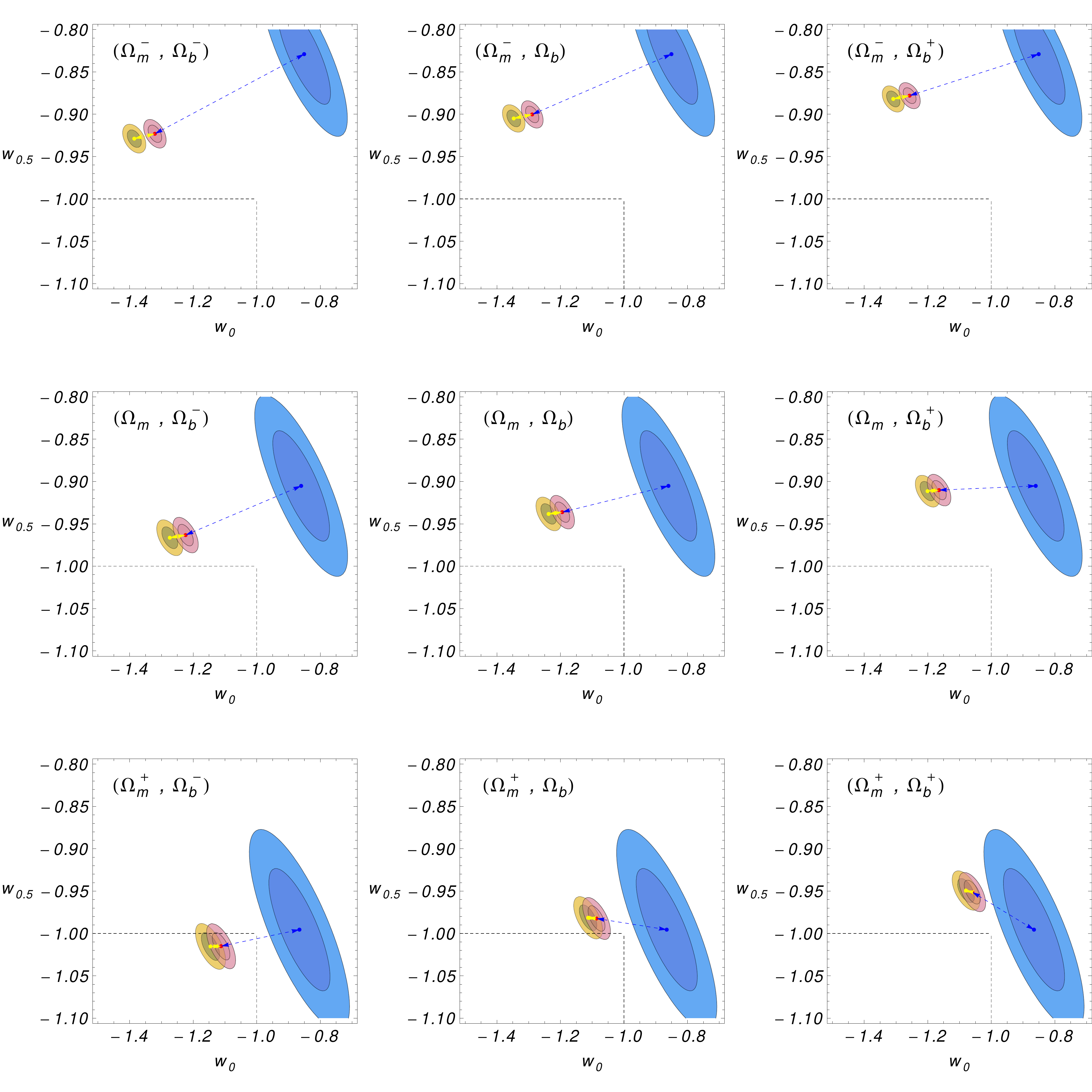}
\caption{Wang model with SNeIa-oriented mock data: 1$\sigma$ and 2$\sigma$ contours in the $w_0$-$w_{0.5}$ parameter space.
In all plots SNeIa are the blue region/point, BAO are the yellow region/point and the combined
SNeIa-BAO are represented by the red region/point. The point where the dashed lines cross indicates the $\Lambda$CDM model. The
arrows indicate the $\sigma$-distances for SNeIa and BAO, with the total $\chi^2$ as measure.}
\label{fig:Wang_SN}
\end{figure*}

\begin{figure*}
\centering
\includegraphics[width=18cm]{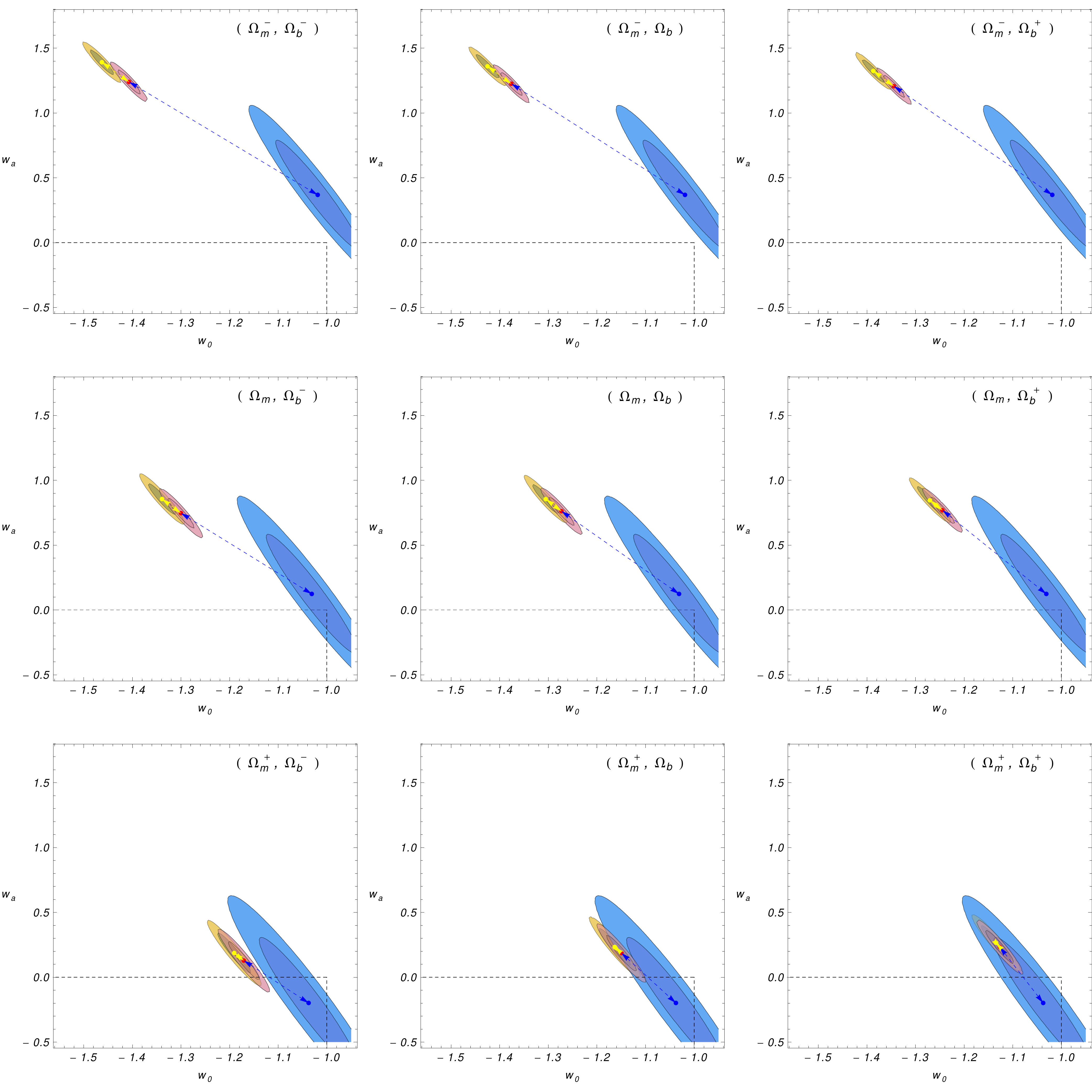}
\caption{CPL model with BAO-oriented mock data: 1$\sigma$ and 2$\sigma$ contours in the $w_0$-$w_a$ parameter space.
In all plots SNeIa are the blue region/point, BAO are the yellow region/point and the combined
SNeIa-BAO are represented by the red region/point. The point where the dashed lines cross indicates the $\Lambda$CDM model. The
arrows indicate the $\sigma$-distances for SNeIa and BAO, with the total $\chi^2$ as measure.}
\label{fig:CPL_BAO}
\end{figure*}

\begin{figure*}
\centering
\includegraphics[width=18cm]{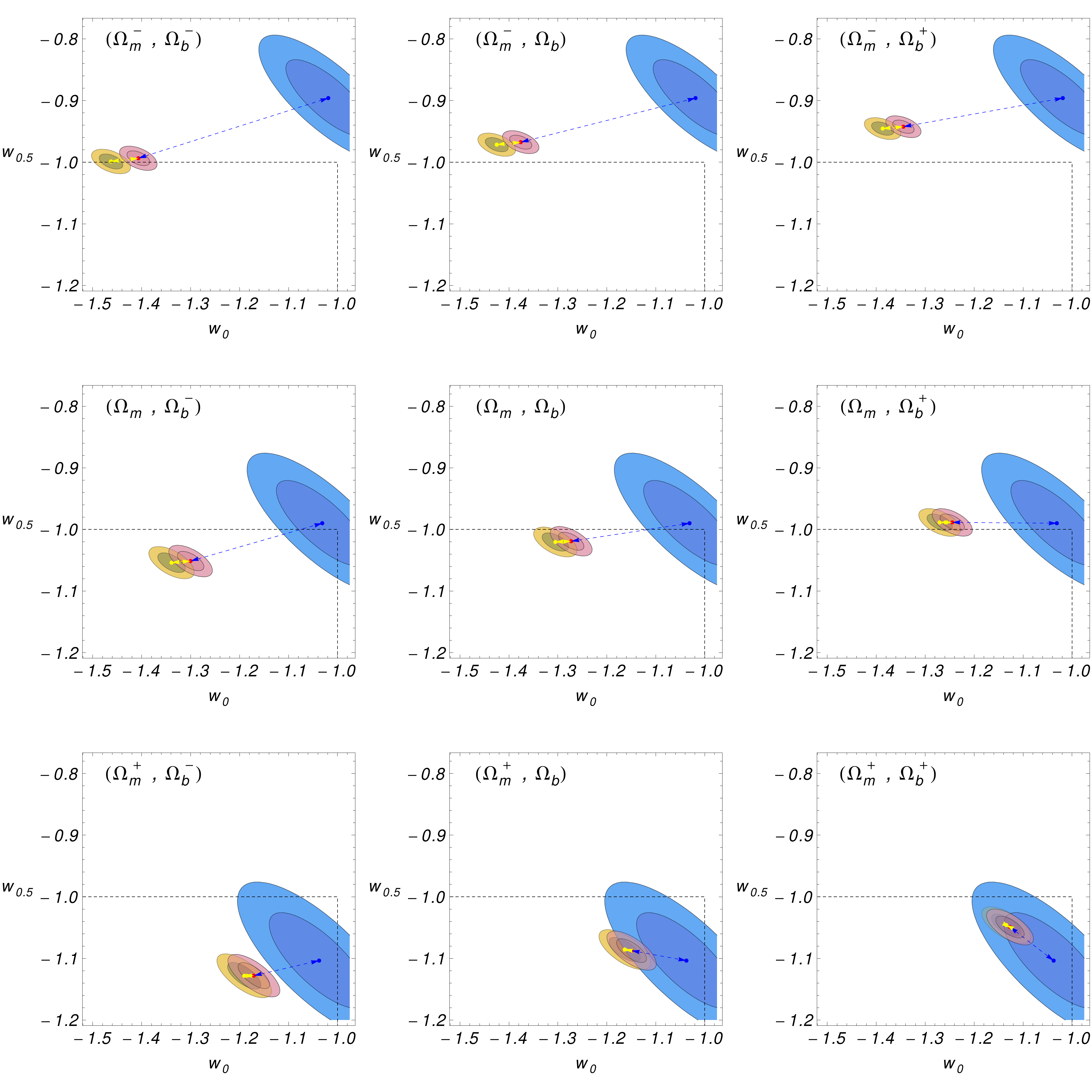}
\caption{CPL model with BAO-oriented mock data: 1$\sigma$ and 2$\sigma$ contours in the $w_0$-$w_a$ parameter space.
In all plots SNeIa are the blue region/point, BAO are the yellow region/point and the combined
SNeIa-BAO are represented by the red region/point. The point where the dashed lines cross indicates the $\Lambda$CDM model. The
arrows indicate the $\sigma$-distances for SNeIa and BAO, with the total $\chi^2$ as measure.}
\label{fig:Wang_BAO}
\end{figure*}

When looking at mock data in Figs.~(\ref{fig:CPL_WMAP})~-~(\ref{fig:Wang_BAO}), we get tension again,
but now the role of SNeIa and BAO is inverted: the tension between SNeIa and SNeIa+BAO
is always larger than $2\sigma$ while the tension between BAO and SNeIa+BAO is much less than $1\sigma$
when $\Omega_m$ has the largest and $\Omega_b$ has the two largest values.\\
Moreover, in this case the joint $\chi^2$ is clearly driven by BAO, exactly the opposite of the real data case. A possible
explanation for this is that, as discussed in the related section, our mock data have been generated using 
different fiducial models, and two of them (the CMB and BAO-oriented case) have a constant value of $w$, below $-1$, even if with large errors).
This choice clearly creates a bias in agreement with previous analyses: the fact that BAO are much keener with
a phantom EoS than SNeIa, combined with the dramatically reduced errors drawn from expectations
produce much smaller credible contours from BAO, very close in position and size to the total ones. In comparison, the statistical role of SNeIa is expected to become subdominant for future data.

On the other hand, we have to note that we have a BAO driven joint $\chi^2$
even in the SNeIa-oriented case, where $w\geq -1$. This may be put down to
the fact that even in this case the smearing effect in the luminosity
data is strong enough to make the data less informative than the BAO ones.
The clear tension between SNeIa and BAO is somewhat surprising given that we have generated both synthetic datasets
using exactly the same fiducial models. The element to have into consideration here is that  the BAO data are divided in transverse and radial
modes. The transverse modes share with SNeIa one fundamental property: they inform us on a quantity
which depends on $w(z)$ through an integral (as can be seen in Eq.~(33)),
and this makes them less sensitive to changes in the EoS
through the so-called \textit{smearing effect} produced by the integration. Moreover the realistic dispersion of data
around the fiducial model induces some dissociation from the underlying model. On the other hand, the radial modes
depend directly on $w(z)$, so the BAO radial component can provide much more refined information.

Assuming the possibility of a substantial smearing effect, and considering the different statistical weights of SNeIa and BAO,
one can wonder if there could possibly be a physical motivation for the tension among BAO and SNeIa, perhaps due  to
something not well established in the BAO or in the SNeIa data (for the SNeIa case, see \cite{Perivolaropoulos09} for a
possible solution).

Finally, looking at the $\sigma-$distances we can also verify that the SNeIa-oriented model seems to be somewhat disfavored
with respect to more phantom models, showing a larger tension (larger values for the $\sigma-$distances) with respect the other
two cases.

\subsection{Dark energy evolution: results}
\label{sec:Prior_EoS}

There is a number of interesting conclusions about the EoS parameters which one can draw from a general overview of our
figures and tables.

\subsubsection{Pivot choice and effect on errors}
First of all, as Wang's model is a reparametrization of the CPL model, it was to be expected that many similarities
between them would be found. In particular, looking at the $\sigma-$distances, we see that
they  match each other perfectly. One can deduce that these two models are different projections
in different parameter subspaces of a more general EoS parameters vector space (i.e. model).
One piece of evidence highlighting the equivalence between between those scenarios is that the
 the estimations of $w_0$ are identical  independently of the  data used.

The main difference lies, obviously, in the EoS pivot  parameter chosen: for
CPL it is $w_a$, whereas Wang takes $w_{0.5}$. Although this cannot be falsified, intuition suggests
Wang's pivot is consistently chosen in a ``redshift sense'', as  it is informing us of a dark energy feature
associated with  a redshift value within the observationally described range. The CPL pivot parameter, on the other hand
is typically very poorly constrained.  Apart from the fact that
 the dark energy evolution in the CPL form is largely redshift independent asymptotically at high redshifts
 ($\lim_{z\to\infty}z/(z+1)=1$), we are questioning ourselves about a physically inaccessible dark energy
 feature, so it is not surprising that for this second dark energy parameter we get exceendingly large errors, as clearly reflected by
 our tables. Wang's remedy  improves the situation substantially; in fact percentual errors on $w_0$ and $w_{0.5}$ are of the same
order of magnitude, whereas for $w_a$ they can even be one order of magnitude larger. This pattern is  common to results coming from
real and
synthetic data.

\subsubsection{Dataset choices/combinations}
In this subsection  we look deeper into the results of our analysis
to gain futher insight on the predictions from different datasets or their combination.
This will allow us to draw more definitive conclusions about evolutionary features, to establish
how convincing those conclusions
are for current data, and to forecast what improvements are expected from future data. Likewise, these results
will throw extra light on the effect of different prior choices (associated with the still considerable uncertainty in the determination
of $\Omega_m$ and $\Omega_b$).

All real data SNeIa cases are  $1\sigma$ compatible with $\Lambda$CDM. However, BAO data favour phantom models
quite manifestly as their best fits are concerned, and the consequence is that for the  SNeIa+CMB combination $\Lambda$CDM typically
  is marginally excluded at $1\sigma$, i.e.
  the case $w_0=-1$ lies outside the $1\sigma$ boundary but very close to it. Arguably, the phantomizing effect of BAO data is
  compensated by their large errors, and possible narrower error bands would yield a more decisive exlusion of $\Lambda$CDM.

Mock data, in contrast with real data, yield a picture in which $\Lambda$CDM is marginally excluded at $1\sigma$ by
SNeIa data, whereas in the BAO and SNeIa+BAO cases the exclusion rate is considerable both because our mock BAO data favour
more phantom models, and because the error bands are narrower.

   On the other hand,  dark energy subdominance at early times, i.e. $w_\infty=w_0 + w_a<0$  is in general not guaranteed, and there are many cases
   (both for real
   and mock data) for which that situation is allowed at $1\sigma$. The main reason for this
   is that $w_a$ is rather poorly constrained. Specifically, for the SNeIa+BAO combination from real data and at the level
    of confidence we have just mentioned one can never guarantee  $w_\infty<0$, whereas this is possible for some mock data cases.
    Another trend
   we observe is that the lowest values of $w_\infty$ correspond to $ \Omega_m=0.260$, the largest value considered.

   A related result is that for SNeIa the larger $\Omega_m$, the lower $w_0$; whereas for BAO the contrary happens and drives
   to the same behaviour for SNeIa+BAO (of course, unless we specify it otherwise, the same pattern is observed for real and mock data).
   This is as well a reflection of the tension between the two datasets. In constrast, increasing $\Omega_m$  induces always a decrease
   in $w_a$. As of $\Omega_b$ we can say that when it decreases, so does $w_0$, but the opposite happens to $w_a$.

In general, if  $\Omega_m$ and $\Omega_b$ are fixed with independent observations, indeterminacies in their values
can affect our conclusions about the current and early values of
   the dark energy parameter $w(z)$. The sensitivity to $\Omega_m$ is quite noticeable, and becomes particularly manifest
   for the $w_a$ parameter in the CPL case, whereas the situation is not so dramatic in Wang's case. Note that even though Wang's parametrization
   did not try to heal correlation between $\Omega_m$ and dark energy parameters,  it comes out as a valuable consequence.
   This parametrization throws another interesting result: for real data and either $\Omega_m<0.260$ or $\Omega_b<0.044$  it can be ascertained at $1\sigma$ that acceleration has increased
   recently ($w$ has become more negative), whereas that situation occurs for mock data in all cases.

\subsection{Main conclusions}

A number of relevant issues about dark energy  have been explored using SNeIa and BAO observational data
within the framework of the most popular representation of its possible evolution: the so called CPL model.
The study is complemented by an in parallel consideration of a Wang's reformulation of the former, as this
is an advantageous alternative which minorates the correlation of parameters. Our main goals were to study
whether there is tension between these two datasets, which are so far two of the most worthy tools
to explore dark energy, and which are anticipated to play an even more preminent role in the future, particularly
due to the spectacular advances expected in BAO astronomical data.

We have confirmed that tension is present in current data, and we have shown that, quite likely, it
will be present in future data. Regarding this aspect of our work, it must be stressed that ours are  the first time simulations of
BAO data from both the radial and transverse directions in the literature.

In addition, we have paid considerable attention to the influence of indeterminacies in the values
of $\Omega_m$ and $\Omega_b$ on the tension issue and other cosmodynamical questions such as
whether dark energy is phantom-like, whether there has been an speed up in the acceleration of the universe,
or whether dark energy could have perhaps been the dominant component at early times. Our main conclusions in this
respect is that for most priors, and for the SNeIa+BAO combination best fits, the Universe is currently phantom-like,
its dark energy EOS parameter has become more negative recently, and was dark matter dominated at early times.

Complementary conclusions are that future BAO data will improve constraints considerably making them far tighter, and
Wang's parametrization is a indeed a very good way to reformulate the CPL model and could  become the dark energy parametrization
of preferred use.


\section{Acknowledgements}
We acknowledge fruitful exchanges with Adam Amara, Bruce Bassett, Narciso Ben\'\i tez, David Parkinson, Karl Glazebrook, Christopher Gordon, Thomas Kitching, Leandros Perivolaropoulos and Anais Rassat.
Celia Escamilla-Rivera is supported by Fundaci\'on Pablo Garc\'ia and FUNDEC, M\'exico.
Irene Sendra holds a PhD FPI fellowship-contract from the Spanish Ministry of Science and Innovation and
Vincenzo Salzano holds a postdoctoral contract from the University of the Basque Country.
Ruth Lazkoz, Vincenzo Salzano and Irene Sendra are supported by the Spanish Ministry of Science and Innovation through research projects FIS2010-15492 and Consolider EPI CSD2010-00064. The four authors are also supported by the Basque Government through the special research action AE-2010-1-31.

\clearpage

{\renewcommand{\tabcolsep}{2.mm}
{\renewcommand{\arraystretch}{1.5}
\begin{table}
\begin{minipage}{\textwidth}
\caption{CPL model and present data. Column.~1: prior choice; columns~2~-~3~-~4: best fit values of $w_0$ and $w_{a}$ for different datasets combinations; columns.~5~-~6~-~7~-~8: $d_\sigma$ calculated as described in Sec.~(\ref{sec:Grid}) with respect to different datasets model.}
\resizebox*{\textwidth}{!}{
\begin{tabular}{cccccccc}
\hline
\multicolumn{1}{c}{$\mathrm{Priors}$} & \multicolumn{3}{c}{$\mathrm{Best \; fit \; parameters \; (w_{0}, w_{a})}$} & \multicolumn{4}{c}{$\sigma-\mathrm{distances}$}\\
\hline \hline
$(\Omega_{m}, \Omega_{b})$ & SN & BAO & SN+BAO & $d_{\sigma}^{\mathrm{SN}}$ & $d_{\sigma}^{\mathrm{BAO}}$ & $d_{\sigma}^{\Lambda}$ & $d_{\sigma}^{\mathrm{BAO-SN}}$\\
\hline
$(0.214,0.0444)$ & $-0.965\pm 0.128$, $0.540\pm 0.736$ & $-1.476\pm 0.713$, $-0.645\pm 6.955$ & $-1.129\pm 0.119$, $1.278\pm 0.697$ & $1.04$ & $>3$ & $1.96$ & $>3$ \\
$(0.237,0.0444)$ & $-0.983\pm 0.137$, $0.367\pm 0.804$ & $-1.171\pm 0.738$, $-2.687\pm 7.255$ & $-1.107\pm 0.128$, $0.917\pm 0.765$ & $0.62$ & $>3$ & $0.88$ & $>3$ \\
$(0.260,0.0444)$ & $-0.999\pm 0.148$, $0.140\pm 0.885$ & $-0.880\pm 0.763$, $-4.666\pm 7.568$ & $-1.084\pm 0.138$, $0.503\pm 0.848$ & $0.30$ & $>3$ & $0.21$ & $>3$ \\
$(0.214,0.0405)$ & $-0.965\pm 0.128$, $0.540\pm 0.736$ & $-1.629\pm 0.725$, $0.046\pm 7.099$  & $-1.153\pm 0.119$, $1.389\pm 0.696$ & $1.27$ & $>3$ & $1.98$ & $>3$ \\
$(0.237,0.0405)$ & $-0.983\pm 0.137$, $0.367\pm 0.804$ & $-1.326\pm 0.752$, $-2.009\pm 7.423$ & $-1.132\pm 0.128$, $1.030\pm 0.764$ & $0.81$ & $>3$ & $0.94$ & $>3$ \\
$(0.260,0.0405)$ & $-0.999\pm 0.148$, $0.140\pm 0.885$ & $-1.036\pm 0.779$, $-4.006\pm 7.762$ & $-1.109\pm 0.138$, $0.620\pm 0.846$ & $0.46$ & $>3$ & $0.34$ & $>3$ \\
$(0.214,0.0366)$ & $-0.965\pm 0.128$, $0.540\pm 0.736$ & $-1.799\pm 0.740$, $0.805\pm 7.272$ & $-1.180\pm 0.119$, $1.514\pm 0.694$ & $1.53$ & $>3$ & $2.03$ & $>3$ \\
$(0.237,0.0366)$ & $-0.983\pm 0.137$, $0.367\pm 0.804$ & $-1.498\pm 0.768$, $-1.269\pm 7.624$ & $-1.159\pm 0.128$, $1.158\pm 0.762$ & $1.05$ & $>3$ & $1.03$ & $>3$ \\
$(0.260,0.0366)$ & $-0.999\pm 0.148$, $0.140\pm 0.885$ & $-1.210\pm 0.798$, $-3.289\pm 7.997$ & $-1.137\pm 0.138$, $0.752\pm 0.844$ & $0.66$ & $>3$ & $0.52$ & $>3$ \\
\hline
\hline\\
\end{tabular}\label{tab:CPL}}
\end{minipage}
\end{table}}}

{\renewcommand{\tabcolsep}{2.mm}
{\renewcommand{\arraystretch}{1.5}
\begin{table}
\begin{minipage}{\textwidth}
\caption{Wang model and present data. Column.~1: prior choice; columns~2~-~3~-~4: best fit values of $w_0$ and $w_{0.5}$ for different datasets combinations; columns.~5~-~6~-~7~-~8: $d_\sigma$ calculated as described in Sec.~(\ref{sec:Grid}) with respect to different datasets model.}
\resizebox*{\textwidth}{!}{
\begin{tabular}{cccccccc}
\hline
\multicolumn{1}{c}{$\mathrm{Priors}$} & \multicolumn{3}{c}{$\mathrm{Best \; fit \; parameters \; (w_{0}, w_{0.5})}$} & \multicolumn{4}{c}{$\sigma-\mathrm{distances}$}\\
\hline \hline
$(\Omega_{m}, \Omega_{b})$ & SN & BAO & SN+BAO & $d_{\sigma}^{\mathrm{SN}}$ & $d_{\sigma}^{\mathrm{BAO}}$ & $d_{\sigma}^{\Lambda\mathrm{CDM}}$ & $d_{\sigma}^{\mathrm{BAO-SN}}$\\
\hline
$(0.214,0.0444)$ & $-0.965\pm 0.128$, $-0.785\pm 0.132$ & $-1.476\pm 0.713$, $-1.691\pm 1.640$ & $-1.129\pm 0.119$, $-0.703\pm 0.128$ & $1.04$ & $>3$ & $1.96$ & $>3$ \\
$(0.237,0.0444)$ & $-0.983\pm 0.137$, $-0.861\pm 0.146$ & $-1.171\pm 0.738$, $-2.067\pm 1.717$ & $-1.107\pm 0.128$, $-0.802\pm 0.143$ & $0.62$ & $>3$ & $0.88$ & $>3$ \\
$(0.260,0.0444)$ & $-0.999\pm 0.148$, $-0.953\pm 0.164$ & $-0.880\pm 0.763$, $-2.436\pm 1.797$ & $-1.084\pm 0.138$, $-0.916\pm 0.161$ & $0.30$ & $>3$ & $0.21$ & $>3$ \\
$(0.214,0.0405)$ & $-0.965\pm 0.128$, $-0.785\pm 0.132$ & $-1.629\pm 0.725$, $-1.613\pm 1.677$ & $-1.153\pm 0.119$, $-0.690\pm 0.127$ & $1.27$ & $>3$ & $1.98$ & $>3$ \\
$(0.237,0.0405)$ & $-0.983\pm 0.137$, $-0.861\pm 0.146$ & $-1.326\pm 0.752$, $-1.996\pm 1.759$ & $-1.132\pm 0.128$, $-0.788\pm 0.142$ & $0.81$ & $>3$ & $0.94$ & $>3$ \\
$(0.260,0.0405)$ & $-0.999\pm 0.148$, $-0.953\pm 0.164$ & $-1.036\pm 0.779$, $-2.372\pm 1.847$ & $-1.109\pm 0.138$, $-0.902\pm 0.160$ & $0.46$ & $>3$ & $0.34$ & $>3$ \\
$(0.214,0.0366)$ & $-0.965\pm 0.128$, $-0.785\pm 0.132$ & $-1.799\pm 0.740$, $-1.530\pm 1.720$ & $-1.180\pm 0.119$, $-0.675\pm 0.127$ & $1.53$ & $>3$ & $2.03$ & $>3$ \\
$(0.237,0.0366)$ & $-0.983\pm 0.137$, $-0.861\pm 0.146$ & $-1.498\pm 0.768$, $-1.921\pm 1.811$ & $-1.159\pm 0.128$, $-0.773\pm 0.142$ & $1.05$ & $>3$ & $1.03$ & $>3$ \\
$(0.260,0.0366)$ & $-0.999\pm 0.148$, $-0.953\pm 0.164$ & $-1.210\pm 0.798$, $-2.306\pm 1.907$ & $-1.137\pm 0.138$, $-0.886\pm 0.160$ & $>3$ & $0.52$ & $>3$ \\
\hline
\hline
\end{tabular}\label{tab:Wang}}
\end{minipage}
\end{table}}}

\clearpage

{\renewcommand{\tabcolsep}{2.mm}
{\renewcommand{\arraystretch}{1.5}
\begin{table}
\begin{minipage}{\textwidth}
\caption{CPL model and mock data. Column.~1: prior choice; columns~2~-~3~-~4: best fit values of $w_0$ and $w_{a}$ for different dataset combinations; columns.~5~-~6~-~7~-~8: $d_\sigma$ calculated as described in Sec.~(\ref{sec:Grid}).}
\resizebox*{\textwidth}{!}{
\begin{tabular}{cccccccc}
\hline
\hline \\
\multicolumn{8}{c}{$\mathrm{CMB}-\mathrm{oriented}$} \\
\hline
\multicolumn{1}{c}{$\mathrm{Priors}$} & \multicolumn{3}{c}{$\mathrm{Best \; fit \; parameters \; (w_{0}, w_{a})}$} & \multicolumn{4}{c}{$\sigma-\mathrm{distances}$}\\
\hline \hline
$(\Omega_{m}, \Omega_{b})$ & SN & BAO & SN+BAO & $d_{\sigma}^{\mathrm{SN}}$ & $d_{\sigma}^{\mathrm{BAO}}$ & $d_{\sigma}^{\Lambda\mathrm{CDM}}$ & $d_{\sigma}^{\mathrm{BAO-SN}}$\\
\hline
$(0.214,0.0444)$ & $-1.058\pm 0.057$, $0.565\pm 0.274$ & $-1.383\pm 0.015$, $1.415\pm 0.055$ & $-1.347\pm 0.014$, $1.314\pm 0.053$ & $>3$ & $2.62$ & $>3$ & $>3$ \\
$(0.237,0.0444)$ & $-1.079\pm 0.061$, $0.366\pm 0.299$ & $-1.276\pm 0.016$, $0.990\pm 0.064$ & $-1.256\pm 0.016$, $0.932\pm 0.063$ & $>3$ & $0.93$ & $>3$ & $>3$ \\
$(0.260,0.0444)$ & $-1.096\pm 0.067$, $0.100\pm 0.328$ & $-1.153\pm 0.019$, $0.489\pm 0.078$ & $-1.152\pm 0.018$, $0.476\pm 0.076$ & $>3$ & $0.06$ & $>3$ & $>3$ \\
$(0.214,0.0405)$ & $-1.058\pm 0.057$, $0.565\pm 0.274$ & $-1.422\pm 0.015$, $1.460\pm 0.057$ & $-1.380\pm 0.014$, $1.343\pm 0.055$ & $>3$ & $>3$   & $>3$ & $>3$ \\
$(0.237,0.0405)$ & $-1.079\pm 0.061$, $0.366\pm 0.299$ & $-1.313\pm 0.017$, $1.015\pm 0.068$ & $-1.287\pm 0.016$, $0.941\pm 0.066$ & $>3$ & $1.44$ & $>3$ & $>3$ \\
$(0.260,0.0405)$ & $-1.096\pm 0.067$, $0.100\pm 0.328$ & $-1.187\pm 0.019$, $0.484\pm 0.083$ & $-1.179\pm 0.019$, $0.456\pm 0.080$ & $>3$ & $0.11$ & $>3$ & $>3$ \\
$(0.214,0.0366)$ & $-1.058\pm 0.057$, $0.565\pm 0.274$ & $-1.462\pm 0.015$, $1.502\pm 0.059$ & $-1.412\pm 0.015$, $1.369\pm 0.058$ & $>3$ & $>3$   & $>3$ & $>3$ \\
$(0.237,0.0366)$ & $-1.079\pm 0.061$, $0.366\pm 0.299$ & $-1.349\pm 0.017$, $1.034\pm 0.071$ & $-1.317\pm 0.017$, $0.944\pm 0.069$ & $>3$ & $1.97$ & $>3$ & $>3$ \\
$(0.260,0.0366)$ & $-1.096\pm 0.067$, $0.100\pm 0.328$ & $-1.219\pm 0.020$, $0.470\pm 0.089$ & $-1.205\pm 0.019$, $0.427\pm 0.086$ & $>3$ & $0.42$ & $>3$ & $>3$ \\
\hline
\hline \\
\multicolumn{8}{c}{$\mathrm{BAO-oriented}$} \\
\hline
\multicolumn{1}{c}{$\mathrm{Priors}$} & \multicolumn{3}{c}{$\mathrm{Best \; fit \; parameters \; (w_{0}, w_{a})}$} & \multicolumn{4}{c}{$\sigma-\mathrm{distances}$}\\
\hline \hline
$(\Omega_{m}, \Omega_{b})$ & SN & BAO & SN+BAO & $d_{\sigma}^{\mathrm{SN}}$ & $d_{\sigma}^{\mathrm{BAO}}$ & $d_{\sigma}^{\Lambda\mathrm{CDM}}$ & $d_{\sigma}^{\mathrm{BAO-SN}}$\\
\hline
$(0.214,0.0444)$ & $-1.019\pm 0.057$, $0.368\pm 0.278$ & $-1.387\pm 0.015$, $1.324\pm 0.058$ & $-1.344\pm 0.015$, $1.206\pm 0.057$ & $>3$ & $>3$ & $>3$ & $>3$ \\
$(0.237,0.0444)$ & $-1.031\pm 0.062$, $0.123\pm 0.304$ & $-1.270\pm 0.017$, $0.845\pm 0.070$ & $-1.245\pm 0.017$, $0.768\pm 0.069$ & $>3$ & $1.36$ & $>3$ & $>3$ \\
$(0.260,0.0444)$ & $-1.037\pm 0.067$, $-0.198\pm 0.333$ & $-1.134\pm 0.020$, $0.265\pm 0.088$ & $-1.127\pm 0.019$, $0.233\pm 0.085$ & $>3$ & $0.10$ & $>3$ & $>3$ \\
$(0.214,0.0405)$ & $-1.019\pm 0.057$, $0.368\pm 0.278$ & $-1.425\pm 0.016$, $1.360\pm 0.061$ & $-1.376\pm 0.015$, $1.225\pm 0.060$ & $>3$ & $>3$   & $>3$ & $>3$ \\
$(0.237,0.0405)$ & $-1.031\pm 0.062$, $0.123\pm 0.304$ & $-1.305\pm 0.018$, $0.854\pm 0.074$ & $-1.273\pm 0.017$, $0.761\pm 0.072$ & $>3$ & $1.89$ & $>3$ & $>3$ \\
$(0.260,0.0405)$ & $-1.037\pm 0.067$, $-0.198\pm 0.333$ & $-1.163\pm 0.021$, $0.233\pm 0.094$ & $-1.149\pm 0.020$, $0.186\pm 0.091$ & $>3$ & $0.36$ & $>3$ & $>3$ \\
$(0.214,0.0366)$ & $-1.019\pm 0.057$, $0.368\pm 0.278$ & $-1.463\pm 0.016$, $1.391\pm 0.063$ & $-1.407\pm 0.016$, $1.239\pm 0.062$ & $>3$ & $>3$   & $>3$ & $>3$ \\
$(0.237,0.0366)$ & $-1.031\pm 0.062$, $0.123\pm 0.304$ & $-1.339\pm 0.019$, $0.855\pm 0.079$ & $-1.300\pm 0.018$, $0.746\pm 0.077$ & $>3$ & $2.43$ & $>3$ & $>3$ \\
$(0.260,0.0366)$ & $-1.037\pm 0.067$, $-0.198\pm 0.333$ & $-1.190\pm 0.022$, $0.186\pm 0.102$ & $-1.171\pm 0.021$, $0.128\pm 0.098$ & $>3$ & $0.79$ & $>3$ & $>3$ \\
\hline
\hline \\
\multicolumn{8}{c}{$\mathrm{SNeIa-oriented}$} \\
\hline
\multicolumn{1}{c}{$\mathrm{Priors}$} & \multicolumn{3}{c}{$\mathrm{Best \; fit \; parameters \; (w_{0}, w_{a})}$} & \multicolumn{4}{c}{$\sigma-\mathrm{distances}$}\\
\hline \hline
$(\Omega_{m}, \Omega_{b})$ & SN & BAO & SN+BAO & $d_{\sigma}^{\mathrm{SN}}$ & $d_{\sigma}^{\mathrm{BAO}}$ & $d_{\sigma}^{\Lambda\mathrm{CDM}}$ & $d_{\sigma}^{\mathrm{BAO-SN}}$\\
\hline
$(0.214,0.0444)$ & $-0.850\pm 0.055$, $0.064\pm 0.266$ & $-1.308\pm 0.014$, $1.280\pm 0.052$ & $-1.257\pm 0.013$, $1.137\pm 0.051$ & $>3$ & $>3$ & $>3$ & $>3$ \\
$(0.237,0.0444)$ & $-0.860\pm 0.059$, $-0.136\pm 0.289$ & $-1.200\pm 0.016$, $0.867\pm 0.061$ & $-1.164\pm 0.015$, $0.762\pm 0.060$ & $>3$ & $2.39$ & $>3$ & $>3$ \\
$(0.260,0.0444)$ & $-0.865\pm 0.064$, $-0.390\pm 0.314$ & $-1.080\pm 0.018$, $0.391\pm 0.072$ & $-1.060\pm 0.017$, $0.325\pm 0.071$ & $>3$ & $0.81$ & $>3$ & $>3$ \\
$(0.214,0.0405)$ & $-0.850\pm 0.055$, $0.064\pm 0.266$ & $-1.347\pm 0.014$, $1.326\pm 0.054$ & $-1.289\pm 0.014$, $1.166\pm 0.053$ & $>3$ & $>3$   & $>3$ & $>3$ \\
$(0.237,0.0405)$ & $-0.860\pm 0.059$, $-0.136\pm 0.289$ & $-1.237\pm 0.016$, $0.897\pm 0.064$ & $-1.194\pm 0.015$, $0.774\pm 0.062$ & $>3$ & $2.93$ & $>3$ & $>3$ \\
$(0.260,0.0405)$ & $-0.865\pm 0.064$, $-0.390\pm 0.314$ & $-1.113\pm 0.018$, $0.395\pm 0.077$ & $-1.086\pm 0.018$, $0.312\pm 0.075$ & $>3$ & $1.30$ & $>3$ & $>3$ \\
$(0.214,0.0366)$ & $-0.850\pm 0.055$, $0.064\pm 0.266$ & $-1.385\pm 0.015$, $1.370\pm 0.056$ & $-1.320\pm 0.014$, $1.192\pm 0.055$ & $>3$ & $>3$   & $>3$ & $>3$ \\
$(0.237,0.0366)$ & $-0.860\pm 0.059$, $-0.136\pm 0.289$ & $-1.273\pm 0.017$, $0.921\pm 0.067$ & $-1.224\pm 0.016$, $0.781\pm 0.065$ & $>3$ & $>3$ & $>3$ & $>3$ \\
$(0.260,0.0366)$ & $-0.865\pm 0.064$, $-0.390\pm 0.314$ & $-1.145\pm 0.019$, $0.390\pm 0.081$ & $-1.112\pm 0.018$, $0.291\pm 0.079$ & $>3$ & $1.82$ & $>3$ & $>3$ \\
\hline
\hline\\
\end{tabular}\label{tab:CPL_mock}}
\end{minipage}
\end{table}}}

\clearpage

{\renewcommand{\tabcolsep}{2.mm}
{\renewcommand{\arraystretch}{1.5}
\begin{table}
\begin{minipage}{\textwidth}
\caption{Wang model and mock data. Column.~1: prior choice; columns~2~-~3~-~4: best fit values of $w_0$ and $w_{0.5}$ for different dataset combinations; columns.~5~-~6~-~7~-~8: $d_\sigma$ calculated as described in Sec.~(\ref{sec:Grid}).}
\resizebox*{\textwidth}{!}{
\begin{tabular}{cccccccc}
\hline
\hline \\
\multicolumn{8}{c}{$\mathrm{CMB}-\mathrm{oriented}$} \\
\hline
\multicolumn{1}{c}{$\mathrm{Priors}$} & \multicolumn{3}{c}{$\mathrm{Best \; fit \; parameters \; (w_{0}, w_{0.5})}$} & \multicolumn{4}{c}{$\sigma-\mathrm{distances}$}\\
\hline \hline
$(\Omega_{m}, \Omega_{b})$ & SN & BAO & SN+BAO & $d_{\sigma}^{\mathrm{SN}}$ & $d_{\sigma}^{\mathrm{BAO}}$ & $d_{\sigma}^{\Lambda\mathrm{CDM}}$ & $d_{\sigma}^{\mathrm{BAO-SN}}$\\
\hline
$(0.214,0.0444)$ & $-1.058\pm 0.057$, $-0.870\pm 0.040$ & $-1.383\pm 0.015$, $-0.911\pm 0.007$ & $-1.347\pm 0.014$, $-0.908\pm 0.007$ & $>3$ & $2.62$ & $>3$ & $>3$ \\
$(0.237,0.0444)$ & $-1.079\pm 0.061$, $-0.957\pm 0.045$ & $-1.276\pm 0.016$, $-0.946\pm 0.008$ & $-1.256\pm 0.016$, $-0.946\pm 0.008$ & $>3$ & $0.93$ & $>3$ & $>3$ \\
$(0.260,0.0444)$ & $-1.096\pm 0.067$, $-1.063\pm 0.050$ & $-1.153\pm 0.019$, $-0.990\pm 0.010$ & $-1.152\pm 0.018$, $-0.993\pm 0.010$ & $>3$ & $0.06$ & $>3$ & $>3$ \\
$(0.214,0.0405)$ & $-1.058\pm 0.057$, $-0.870\pm 0.040$ & $-1.422\pm 0.015$, $-0.936\pm 0.007$ & $-1.380\pm 0.014$, $-0.932\pm 0.007$ & $>3$ & $>3$   & $>3$ & $>3$ \\
$(0.237,0.0405)$ & $-1.079\pm 0.061$, $-0.957\pm 0.045$ & $-1.313\pm 0.017$, $-0.974\pm 0.009$ & $-1.287\pm 0.016$, $-0.973\pm 0.009$ & $>3$ & $1.44$ & $>3$ & $>3$ \\
$(0.260,0.0405)$ & $-1.096\pm 0.067$, $-1.063\pm 0.050$ & $-1.187\pm 0.019$, $-1.025\pm 0.011$ & $-1.179\pm 0.019$, $-1.027\pm 0.011$ & $>3$ & $0.11$ & $>3$ & $>3$ \\
$(0.214,0.0366)$ & $-1.058\pm 0.057$, $-0.870\pm 0.040$ & $-1.462\pm 0.015$, $-0.961\pm 0.007$ & $-1.412\pm 0.015$, $-0.956\pm 0.007$ & $>3$ & $>3$   & $>3$ & $>3$ \\
$(0.237,0.0366)$ & $-1.079\pm 0.061$, $-0.957\pm 0.045$ & $-1.349\pm 0.017$, $-1.005\pm 0.009$ & $-1.317\pm 0.017$, $-1.002\pm 0.009$ & $>3$ & $1.97$ & $>3$ & $>3$ \\
$(0.260,0.0366)$ & $-1.096\pm 0.067$, $-1.063\pm 0.050$ & $-1.219\pm 0.020$, $-1.062\pm 0.012$ & $-1.205\pm 0.019$, $-1.062\pm 0.012$ & $>3$ & $0.42$ & $>3$ & $>3$ \\
\hline
\hline \\
\multicolumn{8}{c}{$\mathrm{BAO-oriented}$} \\
\hline
\multicolumn{1}{c}{$\mathrm{Priors}$} & \multicolumn{3}{c}{$\mathrm{Best \; fit \; parameters \; (w_{0}, w_{0.5})}$} & \multicolumn{4}{c}{$\sigma-\mathrm{distances}$}\\
\hline \hline
$(\Omega_{m}, \Omega_{b})$ & SN & BAO & SN+BAO & $d_{\sigma}^{\mathrm{SN}}$ & $d_{\sigma}^{\mathrm{BAO}}$ & $d_{\sigma}^{\Lambda\mathrm{CDM}}$ & $d_{\sigma}^{\mathrm{BAO-SN}}$\\
\hline
$(0.214,0.0444)$ & $-1.019\pm 0.057$, $-0.896\pm 0.041$ & $-1.387\pm 0.015$, $-0.945\pm 0.007$ & $-1.344\pm 0.015$, $-0.942\pm 0.007$ & $>3$ & $>3$ & $>3$ & $>3$ \\
$(0.237,0.0444)$ & $-1.031\pm 0.062$, $-0.990\pm 0.046$ & $-1.270\pm 0.017$, $-0.989\pm 0.009$ & $-1.245\pm 0.017$, $-0.988\pm 0.009$ & $>3$ & $1.36$ & $>3$ & $>3$ \\
$(0.260,0.0444)$ & $-1.037\pm 0.067$, $-1.104\pm 0.051$ & $-1.134\pm 0.020$, $-1.046\pm 0.012$ & $-1.127\pm 0.019$, $-1.049\pm 0.012$ & $>3$ & $0.10$ & $>3$ & $>3$ \\
$(0.214,0.0405)$ & $-1.019\pm 0.057$, $-0.896\pm 0.041$ & $-1.425\pm 0.016$, $-0.972\pm 0.008$ & $-1.376\pm 0.015$, $-0.968\pm 0.007$ & $>3$ & $>3$   & $>3$ & $>3$ \\
$(0.237,0.0405)$ & $-1.031\pm 0.062$, $-0.990\pm 0.046$ & $-1.305\pm 0.018$, $-1.020\pm 0.010$ & $-1.273\pm 0.017$, $-1.019\pm 0.010$ & $>3$ & $1.89$ & $>3$ & $>3$ \\
$(0.260,0.0405)$ & $-1.037\pm 0.067$, $-0.198\pm 0.333$ & $-1.163\pm 0.021$, $-1.086\pm 0.013$ & $-1.149\pm 0.020$, $-1.087\pm 0.013$ & $>3$ & $0.36$ & $>3$ & $>3$ \\
$(0.214,0.0366)$ & $-1.019\pm 0.057$, $-0.896\pm 0.041$ & $-1.463\pm 0.016$, $-0.999\pm 0.063$ & $-1.407\pm 0.016$, $-0.994\pm 0.008$ & $>3$ & $>3$   & $>3$ & $>3$ \\
$(0.237,0.0366)$ & $-1.031\pm 0.062$, $-0.990\pm 0.046$ & $-1.339\pm 0.019$, $-1.054\pm 0.010$ & $-1.300\pm 0.018$, $-1.051\pm 0.010$ & $>3$ & $2.43$ & $>3$ & $>3$ \\
$(0.260,0.0366)$ & $-1.037\pm 0.067$, $-1.104\pm 0.051$ & $-1.190\pm 0.022$, $-1.128\pm 0.014$ & $-1.171\pm 0.021$, $-1.128\pm 0.014$ & $>3$ & $0.79$ & $>3$ & $>3$ \\
\hline
\hline \\
\multicolumn{8}{c}{$\mathrm{SNeIa-oriented}$} \\
\hline
\multicolumn{1}{c}{$\mathrm{Priors}$} & \multicolumn{3}{c}{$\mathrm{Best \; fit \; parameters \; (w_{0}, w_{0.5})}$} & \multicolumn{4}{c}{$\sigma-\mathrm{distances}$}\\
\hline \hline
$(\Omega_{m}, \Omega_{b})$ & SN & BAO & SN+BAO & $d_{\sigma}^{\mathrm{SN}}$ & $d_{\sigma}^{\mathrm{BAO}}$ & $d_{\sigma}^{\Lambda\mathrm{CDM}}$ & $d_{\sigma}^{\mathrm{BAO-SN}}$\\
\hline
$(0.214,0.0444)$ & $-0.850\pm 0.055$, $-0.829\pm 0.039$ & $-1.308\pm 0.014$, $-0.882\pm 0.006$ & $-1.257\pm 0.013$, $-0.878\pm 0.006$ & $>3$ & $>3$ & $>3$ & $>3$ \\
$(0.237,0.0444)$ & $-0.860\pm 0.059$, $-0.905\pm 0.043$ & $-1.200\pm 0.016$, $-0.911\pm 0.008$ & $-1.164\pm 0.015$, $-0.910\pm 0.007$ & $>3$ & $2.39$ & $>3$ & $>3$ \\
$(0.260,0.0444)$ & $-0.865\pm 0.064$, $-0.995\pm 0.048$ & $-1.080\pm 0.018$, $-0.949\pm 0.009$ & $-1.060\pm 0.017$, $-0.951\pm 0.009$ & $>3$ & $0.81$ & $>3$ & $>3$ \\
$(0.214,0.0405)$ & $-0.850\pm 0.055$, $-0.829\pm 0.039$ & $-1.347\pm 0.014$, $-0.905\pm 0.006$ & $-1.289\pm 0.014$, $-0.900\pm 0.006$ & $>3$ & $>3$   & $>3$ & $>3$ \\
$(0.237,0.0405)$ & $-0.860\pm 0.059$, $-0.905\pm 0.043$ & $-1.237\pm 0.016$, $-0.938\pm 0.008$ & $-1.194\pm 0.015$, $-0.936\pm 0.008$ & $>3$ & $2.93$ & $>3$ & $>3$ \\
$(0.260,0.0405)$ & $-0.865\pm 0.064$, $-0.995\pm 0.048$ & $-1.113\pm 0.018$, $-0.981\pm 0.010$ & $-1.086\pm 0.018$, $-0.982\pm 0.010$ & $>3$ & $1.30$ & $>3$ & $>3$ \\
$(0.214,0.0366)$ & $-0.850\pm 0.055$, $-0.829\pm 0.039$ & $-1.385\pm 0.015$, $-0.929\pm 0.007$ & $-1.320\pm 0.014$, $-0.923\pm 0.007$ & $>3$ & $>3$   & $>3$ & $>3$ \\
$(0.237,0.0366)$ & $-0.860\pm 0.059$, $-0.905\pm 0.043$ & $-1.273\pm 0.017$, $-0.966\pm 0.009$ & $-1.224\pm 0.016$, $-0.963\pm 0.008$ & $>3$ & $>3$ & $>3$ & $>3$ \\
$(0.260,0.0366)$ & $-0.865\pm 0.064$, $-0.995\pm 0.048$ & $-1.145\pm 0.019$, $-1.015\pm 0.011$ & $-1.112\pm 0.018$, $-1.015\pm 0.011$ & $>3$ & $1.82$ & $>3$ & $>3$ \\
\hline
\hline\\
\end{tabular}}\label{tab:Wang_mock}
\end{minipage}
\end{table}}}

\vfill
\end{document}